\documentclass[iop,onecolumn,onecolappendix,numberedappendix]{emulateapj}

\usepackage{amsmath,amsthm}
\usepackage{booktabs}
\usepackage{threeparttable}
\usepackage{tabularx}
\usepackage{xcolor}

\usepackage{times}
\usepackage{graphicx}
\usepackage{comment}

\usepackage{lineno}

\newcommand{\Lx}{L_{\rm X}}
\newcommand{\Tx}{T_{\rm  X}}
\newcommand{\Tc}{T_{\rm  c}}
\newcommand{\Mst}{M_*}
\newcommand{\rhost}{\rho_*}
\newcommand{\rst}{r_*}
\newcommand{\etas}{\eta_*}
\newcommand{\etag}{\eta_{\rm g}}
\newcommand{\qs}{q_*}
\newcommand{\Reff}{R_{\rm e}}

\newcommand{\RdHI}{R_{\rm d HI}}
\newcommand{\Rds}{R_{\rm d*}}
\newcommand{\Rhalf}{R_{1/2}}
\newcommand{\Sigds}{\Sigma_{\rm d *}}

\newcommand{\Mdstar}{M_{\rm d *}}
\newcommand{\MdHI}{M_{\rm d HI}}
\newcommand{\SdHI}{\langle\Sigma_{\rm d HI}\rangle}

\newcommand{\phids}{\phi_{\rm d *}}

\newcommand{\MR}{{\cal R}}
\newcommand{\MRm}{\MR_{\rm m}}
\newcommand{\Mgal}{M_{\rm g}}
\newcommand{\rgal}{r_{\rm g}}
\newcommand{\rhogal}{\rho_{\rm g}}
\newcommand{\qg}{q_{\rm g}}
\newcommand{\Mbh}{M_{\rm BH}}
\newcommand{\dMbh}{\dot M_{\rm BH}}
\newcommand{\dMedd}{\dot M_{\rm Edd}}
\newcommand{\Mout}{M_{\rm out}}
\newcommand{\Mhot}{M_{\rm hot}}
\newcommand{\vh}{v_{\rm h}}
\newcommand{\rh}{r_{\rm h}}
\newcommand{\csih}{\xi_{\rm h}}
\newcommand{\rhoh}{\rho_{\rm h}}
\newcommand{\Mh}{M_{\rm h}}
\newcommand{\phigal}{\phi_{\rm g}}
\newcommand{\phih}{\phi_{\rm h}}

\newcommand{\vbh}{v_{\rm BH}}
\newcommand{\vgal}{v_{\rm g}}
\newcommand{\vcirc}{v_{\rm c}}
\newcommand{\Rec}{\langle R_{\rm e} \rangle}
\newcommand{\Msol}{{\rm M}_\odot}
\newcommand{\kpc}{{\rm kpc}}
\newcommand{\kms}{{\rm km}\;{\rm s}^{-1}}

\newcommand{\MDM}{M_{\rm DM}}
\newcommand{\LB}{L_{\rm B}}
\newcommand{\LBsol}{L_{\rm B,\odot}}

\newcommand{\Dels}{\Delta_*}

\newcommand{\sigmaBH}{\sigma_{*{\rm BH}}}
\newcommand{\sigmag}{\sigma_{*{\rm g}}}
\newcommand{\sigmas}{\sigma_*}
\newcommand{\sigmass}{\sigma_{**}}

\newcommand{\DeltaBH}{\Delta_{*{\rm BH}}}
\newcommand{\Deltag}{\Delta_{*{\rm g}}}
\newcommand{\Deltas}{\Delta_*}
\newcommand{\Deltass}{\Delta_{**}}

\newcommand{\phiT}{\phi_{\rm tot}}
\newcommand{\phiBH}{\phi_{\rm BH}}
\newcommand{\phis}{\phi_*}
\newcommand{\phin}{\phi_{\rm n}}

\newcommand{\tphisz}{\tilde{\phi}_{*0}}
\newcommand{\tphisu}{\tilde{\phi}_{*1}}
\newcommand{\tphisd}{\tilde{\phi}_{*2}}
\newcommand{\tphisi}{\tilde{\phi}_{*i}}

\newcommand{\vphib}{\overline{v_\varphi}}
\newcommand{\sigmaphi}{\sigma_\varphi}

\newcommand{\Hcsi}{{\cal H}}

\newcommand{\ms}{m_*}
\newcommand{\mg}{m_{\rm g}}

\newcommand{\tz}{t_0}
\newcommand{\Macc}{M_{\rm acc}}
\newcommand{\rt}{r_{\rm t}}

\newcommand{\ka}{k_{\rm a}}
\newcommand{\ke}{k_{\rm e}}

\newcommand{\Rin}{R_{\rm in}}
\newcommand{\Rfin}{R_{\rm fin}}

\newcommand{\epsw}{\epsilon_{\rm w}}
\newcommand{\epswM}{\epsilon_{\rm w}^{\rm M}}

\slugcomment{Revised, May 10, 2022}

\begin{document}

\shortauthors{L. Ciotti, J.P. Ostriker, Z. Gan, B.X. Jiang, S. Pellegrini,
  C. Caravita, A. Mancino} 
\shorttitle{A parameter-space exploration of X-ray halos of rotating ETGs with AGN feedback}


\title{A Parameter Space Exploration of High Resolution Numerically Evolved Early Type\\ Galaxies 
Including AGN Feedback and Accurate Dynamical Treatment of Stellar Orbits}

\author{Luca Ciotti}
\affiliation{Department of Physics and Astronomy, University of  Bologna, via Gobetti 93/2, I-40129 Bologna, Italy}

\author {Jeremiah P. Ostriker}
\affiliation{Department of Astronomy, Columbia University, 550 West 120th St, New York, NY 10027, USA\\
Department of Astrophysical Sciences, Princeton University, Princeton, 
NJ 08544, USA}

\author{Zhaoming Gan}
\affiliation{New Mexico Consortium, Los Alamos, NM 87544, USA \\
Department of Astronomy, Columbia University, 550 W, 120th Street, New York, NY 10027, USA}

\author{Brian Xing Jiang}
\affiliation{Department of Astronomy, Columbia University, 550 
  W. 120th Street, New York, NY 10027, USA}

\author{Silvia Pellegrini, Caterina Caravita, Antonio Mancino}
\affiliation{Department of Physics and Astronomy, University of
  Bologna, via Gobetti 93/2, I-40129 Bologna, Italy\\
  INAF-Osservatorio di Astrofisica e Scienza dello Spazio di Bologna, Via Gobetti 93/3, Bologna I-40129, Italy}

\begin{abstract}

  An extensive exploration of the model parameter space of
  axisymmetric Early-Type Galaxies (ETGs) hosting a central
  supermassive Black Hole (SMBH) is conducted by means of high
  resolution hydrodynamical simulations performed with our code
  MACER. Global properties such as 1) total SMBH accreted mass, 2)
  final X-ray luminosity and temperature of the X-ray emitting halos,
  3) total amount of new stars formed from the cooling gas, 4) total
  ejected mass in form of supernovae and AGN feedback induced galactic
  winds, are obtained as a function of galaxy structure and internal
  dynamics. In addition to the galactic dark matter halo, the model
  galaxies are also embedded in a group/cluster dark matter halo;
  finally cosmological accretion is also included, with amount and
  time dependence derived from cosmological simulations. Angular
  momentum conservation leads to the formation of cold HI disks; these
  disks further evolve under the action of star formation induced by
  disk instabilities, of the associated mass discharge onto the
  central SMBH, and of the consequent AGN feedback.  At the end of the
  simulations, the hot (metal enriched) gas mass is roughly $10\%$ the
  mass in the old stars, with twice as much having been ejected into
  the intergalactic medium. The cold gas disks are a $\approx$ kpc in
  size, and the metal rich new stars are in $0.1$ kpc disks.  The
  masses of cold gas and new stars are roughly $0.1\%$ the mass of the
  old stars. Overall, the final systems appear to reproduce quite
  successfully the main global properties of real ETGs.

\end{abstract}

\keywords{galaxies: elliptical and lenticular, cD -- 
galaxies: evolution --
quasars: supermassive black holes --
X-rays: galaxies -- 
X-rays: ISM}

\section{Introduction}

Numerous observational, numerical, and theoretical studies show that
in Early Type Galaxies (hereafter ETGs), the evolution of their hot
X-ray emitting atmospheres (e.g., Kim et al. 2019, Babyk et al. 2018)
is determined by the complex interplay between the ISM (produced by
stellar mass losses and cosmological accretion from group/cluster
environment), and the internal structure and dynamics of the host
galaxies, the SNIa's heating, the central SMBH AGN feedback effects
(see for reviews Kim and Pellegrini 2012, Mathews and Brighenti 2003,
Werner et al. 2019).  Over the years, increasingly detailed and
realistic simulations have been developed and performed by several
groups, with specific focus on the several aspects of the problem. The
improvements on the input physics can be broadly summarized in four
large categories: 1) galaxy structure and internal dynamics (e.g.,
shape and density profiles of stars and dark matter, velocity
dispersion and rotational fields of the stellar component), 2) physics
of the ISM (cooling and heating mechanisms, evolution of the dust and
metals content of the ISM, star formation processes, instabilities),
3) central SMBH accretion and associated AGN feedback (radiative and
mechanical feedback and its dependence on the local ISM properties,
radiative transfer, cosmic-ray acceleration, 4) cluster/group
confining and accretion effects.  Recent studies of our group with the
hydro code MACER (Gan et al. 2019a,b, Gan et al. 2020, hereafter
G19a,b and G20, respectively, and references therein), focused mainly
on point 3), with exploratory investigations of points 1) and 2), in
particular concerning the effect of galaxy shape and rotation on the
SMBH accretion, gas cooling, and star formation. These high-resolution
axisymmetric hydrodynamical simulations have inner boundaries ranging
from 2.5 pc to 20 pc to resolve the Bondi radius. And, while only
performed in 2D, they greatly exceed the spatial resolution available
in most cosmological simulations.

In particular, effects of galaxy shape and rotation seem to deserve
special attention, motivated by observational and theoretical
arguments. In fact rotating and flat ETGs are observed to host (albeit
with the usual non-negligible scatter in their properties)
systematically fainter and cooler X-ray emitting halos than ETGs of
same optical luminosity but of rounder shape and with less ordered
rotation in the stellar population (see, e.g., Eskridge et al. 1995,
Juranova et al. 2020, Kim and Fabbiano 2015, Pellegrini et al. 1997,
Sarzi 2013).  Preliminary simulations, conducted with a different 2D
code in cylindrical coordinates (Posacki et al. 2013, Negri et
al. 2014a,b, Negri et al. 2015) reassuringly showed that in fact
rotation can be  effective in enhancing ISM instabilities and
leading to the formation of cold gaseous rotating disks, with
substantial reduction of X-ray luminosity, and lower emission
temperatures of the ISM (see also Brighenti and Mathews 1996, 1997;
D'Ercole and Ciotti 1998, and references therein).  These simulations,
while modeling star formation in the equatorial gaseous disk by using
a simple, physically based recipe for star formation, lacked however
the modeling of angular momentum transport, with the consequent
inability to properly model SMBH accretion, so that in these
preliminary simulations AGN feedback was not activated.  A step
forward in the modeling, also including physically appropriate AGN
feedback, confirming the main results of these preliminary
investigations, was done in a series of subsequent papers (Ciotti et
al. 2017, Pellegrini et al. 2018, Yoon et al. 2018, G20), by using
galaxy models of increasing realism.

For what concerns the numerical modeling of gas flows in ETGs, two
complementary approaches can be devised, each of them with its merits
and limitations. In the first, one focuses on some specific, well
observed galaxy, and attempts to reproduce in detail the observed
features (in particular, the X-ray surface brightness profile and the
the temperature profile of the ISM), to test the implemented physical
assumptions. In the second, one instead considers a large set of
galaxy models, spanning the range of observed galaxy properties,
aiming at reproducing the observed trends of global properties, such
as the ISM total X-ray luminosity and emission-weighted temperature,
the final SMBH masses, the duty-cycle of the AGN, and so on. Of
course, in the first approach one can use well taylored galaxy models,
but the unavoidable shortcoming is that one does not have information
on the specific time at which the real system is observed, a problem
somewhat aggravated by the empirical (and significant) differences
from system-to-system: in practice, also when modeling a well observed
galaxy, from the point of view of the simulations, one is forced to
interpret the results in some time-averaged way. In the second
approach, one cannot expect to reproduce in great detail a single
object, however global trends (presumably quite independent of very
specific physical assumptions, and averaged over the large number of
models) may be reproduced, thus hopefully deriving information useful
to build a consistent ``big picture'' of the different physical
mechanisms involved in the evolution of the ISM, and in the AGN
feedback activity.  Clearly, the exploration of the parameter space
can be very time-expensive, in particular if high spatial and temporal
resolution is adopted (as required for a proper numerical modeling).

In this paper we take advantage of the latest version of our
high-resolution MACER code, improved in particular on the physical
treatment of feedback, on the effects of rotation of the stellar
component on SMBH accretion, and on star formation and disk
instabilities, and we focus on the second approach, by using realistic
dynamical models for the host galaxies. In particular, an extensive
exploration of the model parameter space is conducted. Properties such
as 1) total SMBH accreted mass, 2) final X-ray luminosity and
temperature of the X-ray emitting halos, 3) total amount of new stars,
4) total ejected mass are obtained, as a function of galaxy structure
(stellar and DM amount and distribution, galaxy flattening), and
internal dynamics (amount of ordered rotation). A group/cluster DM
halo is also added, providing an important confining effect, and
finally cosmological accretion is also included, in accordance with
the results of cosmological simulations (at this stage however a major
omission is the neglect of the accretion of satellite galaxies). We
also consider the time change of the stellar velocity dispersion and
rotational velocity fields, due to mass loss of the stellar
populations, and to the mass growth of the central SMBH. The code used
has been developed by Ciotti \& Ostriker and collaborators (2001,
2007, 2011), with recent major additions described in G19a,b and G20,
to allow for the inclusion of a suite of chemical elements, and the
study of dust production and destruction.

The paper is organized as follows: in Section 2 we describe the galaxy
models adopted for the simulations, and in Section 3 we present the major
upgrades in the input physics. Section 4 is dedicated to present the
main results, while in Section 5 we discuss the results and present the
conclusions, together with a list of important improvements that we
are currently developing.

\section{The galaxy models}
\label{mod}

A major ingredient for the hydrodynamical simulations of galactic gas
flows is represented by the galaxy models hosting the flows. In fact,
the models are needed in order to assign the gravitational field of
the host galaxies, and the spatial and temporal distribution of the
gas source terms (mass, momentum, and energy). In turn, the momentum
and energy terms require the specification of the galaxy internal
dynamics.  Over the years, more and more realistic (and numerically
tractable) models have been developed and employed in the simulations.

The galaxy models adopted here are an extension of the models already
used in G19a,b and G20, and are based on the JJe dynamical models
(Ciotti et al. 2021, hereafter CMPZ21). Here we recall their main
structural and dynamical properties relevant for the hydrodynamical
simulations.  The stellar density distribution is described by an
oblate ellipsoidal Jaffe (1983) model of total mass $\Mst$,
scale-length $\rst$, and axial ratio $0<\qs\leq 1$, so that its
density profile is given by
\begin{equation}
      \rhost(\ms)= 
      {\Mst\over 4\pi \qs\rst^3\ms^2 (1+\ms)^2},
      \quad 
      \ms^2 \equiv {R^2\over\rst^2}+{z^2\over\qs^2\rst^2}.
\label{eq:rhos}
\end{equation}
It is useful to introduce the flattening parameter $\etas$, related to
the axial ratio as $\etas=1-\qs$, so that $\etas=0$ corresponds to a
spherical stellar distribution. The circularized
effective radius $\Rec$ of an ellipsoidal stellar system observed edge-on is
related to the effective radius $\Reff$ of the same model in the
spherical limit (or when observed face-on) by the identity
\begin{equation}
\Rec=\sqrt{\qs}\Reff,
\label{eq:Reff}
\end{equation}
moreover, in the edge-on projection of an ellipsoidal system, the
isophotal flattening coincide with the intrinsic flattening. As is
well known, the projected density profile of the Jaffe model is
remarkably similar to the de Vaucouleurs $R^{1/4}$ law over a quite
large radial range, and in the spherical case $\Reff\simeq 3\rst/4$,
so that we can use Equation (\ref{eq:Reff}) to determine the scale
$\rst$ for our models once $\qs$ and $\Rec$ are fixed by observations.

In JJe models the stellar distribution is embedded in a {\it galactic}
dark matter (hereafter DM) halo, so that the {\it total} (stellar plus
DM) galaxy density distribution is again described by a Jaffe
ellipsoidal distribution of total mass $\Mgal=\MR\Mst$, axial ratio
$\qg$, and scale length $\rgal=\xi\rst$:
\begin{equation}
      \rhogal(\mg)= 
      {\Mst\MR\xi\over 4\pi\rst^3\qg\mg^2 (\xi+\mg)^2},
      \quad 
      \mg^2 \equiv {R^2\over\rst^2}+{z^2\over\qg^2\rst^2};
\label{eq:rhog}
\end{equation}
in the present models we always assume the natural choice $\xi\geq 1$.
As in our previous papers in this series (G19a,b, G20), for simplicity
we restrict to the case of spherically symmetric $\rhogal$, i.e., we
set $\qg=1$ in Equation (\ref{eq:rhog}). The approximation is quite
acceptable for moderately flattened galaxies (as the isopotential
surfaces are in general rounder than the associated mass density),
with the additional advantage of a simple expression for the galaxy
gravitational field, and of explicit expressions for the solutions of
the Jeans equations, of easy implementation in the hydrodynamical code
(see Section 2.1).  In the spherical limit, the total galaxy mass
contained in the sphere of radius $r$, and the galaxy potential, are
given by
\begin{equation}
      \Mgal(r)= 
      {\Mst\MR s\over\xi +s},
      \qquad
            \phigal(r)=
            {G\Mst\MR\over \rst\xi}\ln {s\over\xi+s},
            \qquad
      s\equiv {r\over\rst}.
\label{eq:phigal}
\end{equation}
Since in JJe models $\rhogal$ and $\rhost$ are assigned, a condition
for the positivity of the galaxy DM halo density distribution
$\rho_{\rm DM}=\rhogal -\rhost$ is needed.  From Equation (13) in
CMPZ21, imposing $\xi\geq 1$ and $\qg =1$, the positivity condition
reduces to
\begin{equation}
  \MR\geq\MRm ={\xi\over 1-\etas},
  \label{eq:minhal}
\end{equation}
A model with $\MR=\MRm$ is called {\it minimum halo} model, and it can
be shown that in this case $\rho_{\rm DM}$ is well described by the
NFW profile over a large radial range (see Ciotti \& Ziaee Lorzad
2018; Ciotti, Mancino and Pellegrini 2019, CMPZ21), for this reason
{\it in the simulations we set the initial galaxy parameters to the
  minimum-halo case}. Notice that the total (stars plus DM) galaxy
density profile$\rhogal$ in Equation (\ref{eq:rhog}) is proportional
to $r^{-2}$ inside $\rgal$: this property is one of the motivations
behind the construction of JJe models, as different theoretical and
observational findings support this assumption over a large radial
range (e.g., see among others, Wang et al. 2019, 2020; Li, Shu and
Wang 2018; Cappellari et al. 2015; Poci, Cappellari and McDermid 2017;
Lyskova, Churazov and Naab 2018; Auger et al. 2010; Barnab\`e et
al. 2011; Koopmans et al. 2009; Gavazzi et al. 2007;
Serra et al. 2016; Bellstedt et al. 2018).

In order to take into account the effects of a group/cluster DM halo
on the gas flows, we also consider the gravitational field produced by
a spherically symmetric quasi-isothermal DM halo of asymptotic circular
velocity $\vh$ and scale-length $\rh=\csih\rst$
\begin{equation}
      \rhoh(r)=
      {\vh^2\over 4\pi G\rst^2(\csih^2 +s^2)},
\label{eq:rhoh}
    \end{equation}
\begin{equation}
      \Mh(r)=
      {\vh^2\rst\over G}\,\left(s-\csih\arctan {s\over\csih}\right),\qquad
      \phih(r)=
      \vh^2\,\left(\ln{\sqrt{ 1+s^2/\csih^2}\over {\rm e}}+{\csih\over
          s}\arctan {s\over\csih}\right). 
\label{eq:phih}
\end{equation}
Notice that in Equations (\ref{eq:phigal}) and (\ref{eq:phih}) we
fixed $\phigal(\infty)=0=\phih(0)$.  In the simulations we consider
models with $\csih \gg 1$, and so, as we will see in the next
Section, the group/cluster DM component does not alter significantly
the internal dynamics of the models (see Section 2.1).

The stellar mass $\langle\Mst\rangle (r)$ contained in a sphere of
radius $r$ centered in the origin is easily computed in the homeoidal
expansion approximation, and from Equations (15)-(16) in CMPZ21 we
have
\begin{equation}
      \langle\Mst\rangle(r) 
      ={\Mst s\over 1+s}\left[1+{\etas\over3(1+s)}\right], 
\label{eq:mstcirc}
\end{equation}
so that the total DM mass (galactic plus group/cluster) inside the
same sphere is
\begin{equation}
      \MDM(r)={\Mst\MR s\over
        \xi+s}-\langle\Mst\rangle(r)+\Mh(r).
\label{eq:mdmcirc}
\end{equation}
For the three families of models in Table \ref{models}, it follows
that $\MDM(r)/\Mgal(r)\simeq 52\%$ at $r=\Rec$, and $\simeq 64\%$ at
$r=2\Rec$ for $\etas=0.3$, and for reference $\simeq 39\%$ at
$r=\Reff$, and $\simeq 55\%$ at $r=2\Reff$, for $\etas=0$.

\begin{figure}
  \hskip 0.1truecm 
  \includegraphics[width=0.5\linewidth, keepaspectratio]{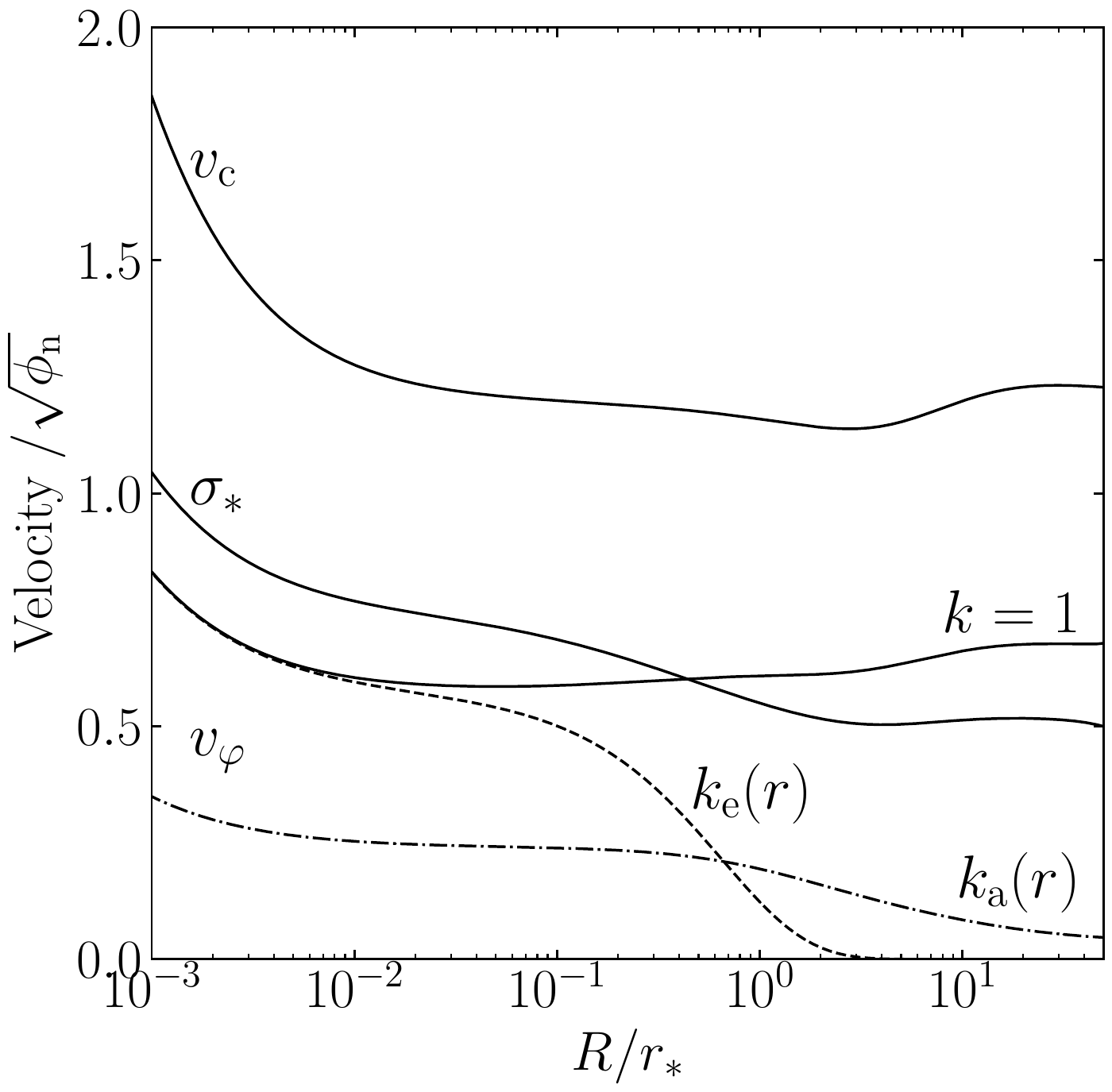}
  \includegraphics[width=0.5\linewidth, keepaspectratio]{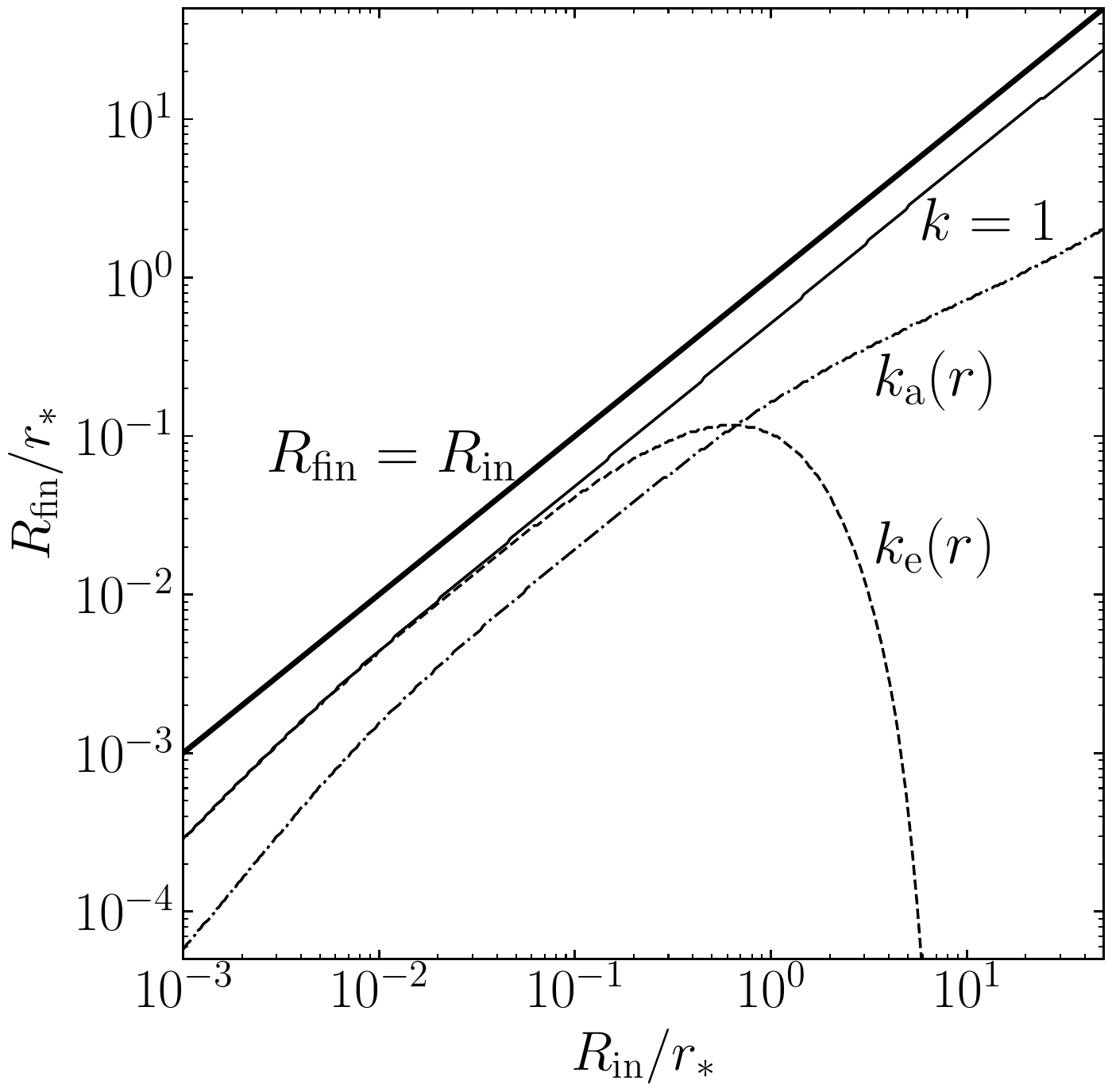}  
  \caption{Left: radial profiles of the stellar velocities in the
    equatorial plane in units of
    $\sqrt{\phin}\equiv\sqrt{G\Mst/\rst}$ (the numerical values of
    this normalization scale are given in Column (7) of
    Table~\ref{models} for the different model families). From top to
    bottom, the circular velocity $\vcirc$ in Equation
    (\ref{eq:vcircTOTAL}), the stellar vertical velocity dispersion
    component $\sigmas$ in Equation (A1), and the three different
    azimuthal streaming velocities $\vphib$ obtained from the first of
    Equations (\ref{eq:vphis}), respectively for the isotropic rotator
    (solid, $k=1$), and for the exponentially declining (dashed,
    $k=\ke$) and asymptotically flat (dot-dashed, $k=\ka$) Satoh
    decompositions in Equation (\ref{eq:satohkr}). The effect of the
    central SMBH is clearly visibile in the innermost regions; notice
    also that in the exponential decomposition the stars rotate faster
    in the inner regions (as the isotropic rotator) than in the
    asymptotically flat decomposition, while rotation is the lowest in
    the outher galactic regions. Right: final circularization radius
    $\Rfin$ for the gas infalling on the equatorial plane at the
    radius $\Rin$, under the assumption of angular momentum
    conservation discussed in Section 4.2 (see in particular Equation
    \ref{eq:djzdt}).  The solid line refers to isotropic rotators, the
    dot-dashed line to the asymptotically flat decomposition, and the
    dashed line to the exponential decomposition. The heavy solid line
    marks the locus of $R_{\rm f in} = \Rin$.}
  \label{f1}
\end{figure}
\vskip 0.8truecm 

Finally, a SMBH of initial mass $\Mbh=\mu\Mst$ (with an initial value
of $\mu\simeq 10^{-3}$, half of the currently observationally
estimated value) is added at the center of the galaxy, with
\begin{equation}
      \phiBH(r)= -{G\Mst\mu\over r}.
\label{eq:phiBH}
\end{equation}

\subsection{Internal Dynamics}\label{sec:internal_dynamics}

The internal dynamics of the galaxy models, i.e., their velocity
dispersion and ordered rotation fields, are important ingredients of
the problem, as they determine the momentum and kinetic energy sources
associated with stellar mass losses that enter the hydrodynamical
equations. The kinematical fields are obtained by solving the Jeans
Equations for the density $\rhost$, under the assumption of a
two-integral phase-space distribution function; here we just recall
the main properties relevant for the setup of the simulations (see
CMPZ21, for a complete description of the models).  In particular, the
Jeans equations for the stellar component are solved in homeoidal
approximation, so that the solution can be expressed in fully
analytical form (see Appendix A). This fact not only allows for a
simple numerical implementation (G19a), but it also allows to follow
the secular changes of the gravitational and kinematical fields due to
the stellar mass losses and the mass growth of the central SMBH, just
by imposing the required time dependence on the structural parameters
(see Appendix B).  We also consider the effects on the gas flows of
the (time dependent) gravitational field associated with the formation
of the stellar disk in the equatorial plane (Section 3), and with the
gravitational field of a group/cluster DM halo; for simplicity,
instead, we do not consider their effects on the stellar kinematical
field, so the formulae in the Appendices give the kinematical field
produced by the total mass distribution (disk excluded) and the
central SMBH.

\renewcommand\arraystretch{1.4}
\begin{table*}
\centering 
\caption{Structural properties of the three families of models}
\vspace{2mm}
\begin{threeparttable}
\begin{tabularx}{\textwidth}{c@{\extracolsep{\fill}}ccccccccccccc}
\toprule 
  Model family   &  $\LB$       &  $\Mst$          & $\rst$   & $\Rec$   &$\vcirc(0)$ & $\sigmas(0)$& $\sqrt{G\Mst/\rst}$ & $\vh$\\
     & $(10^{11}\,\LBsol)$ &$(10^{11}\,\Msol)$& $(\kpc)$&$(\kpc)$ &  $(\kms)$  &  $(\kms)$  &  $(\kms)$ & $(\kms)$   \\
                             & (1)         &      (2)              & (3)         &      (4)     &  (5)            &   (6)              &      (7)     &  (8)   \\ 
 \midrule                                                 
  LM  & $0.32$ & $1.54$ & $7.33$     & $4.57$  & $360$  & $223$  & $301$  & $360$  \\
  \midrule 
  MM  & $0.65$ & $3.35$  & $11.29$  & $7.04$  & $427$  & $265$ &$357$ & $427$ \\                                                                                                  
 \midrule 
  HM & $1.38$ & $7.80$ & $18.94$    & $11.80$  & $504$  &  $312$& $421$ & $504$ \\
  \bottomrule 
\end{tabularx}
\vspace{1.5mm}
\begin{tablenotes}[para,flushleft]
  \footnotesize For the family name on the left, each column gives:
  (1) the galaxy luminosity in the $B$-band, (2) the initial stellar
  mass, (3) the scale-length of the stellar distribution (Equation
  \ref{eq:rhos}), (4) the edge-on circularized effective radius
  (Equation \ref{eq:Reff}), (5) the galaxy central circular velocity
  (in absence of the SMBH and in the minimum halo case, Equation
  \ref{eq:vcirc_zero}), (6) the stellar central velocity dispersion
  (in absence of the SMBH, and in the minimum halo case,
  Equation \ref{eq:sigmazz}), (7) the velocity scale of the models,
  and (8) the asymptotic circular velocity of the quasi-isothermal DM
  halo (Equation \ref{eq:rhoh}), fixed to coincide with
    $\vgal(0)$.  For all models, the flattening of the stellar
  distribution in Equation (\ref{eq:rhos}) is $\etas=0.3$, the initial
  SMBH-to-stellar mass ratio is $\mu=\Mbh/\Mst = 0.001$, the
  parameters $\xi$ and $\MR$ characterizing the total galaxy density
  in Equation (\ref{eq:rhog}) are $\xi =12.6$ and $\MR=18$,
  corresponding to a minimum-halo model from Equation
  (\ref{eq:minhal}), and the scale-length of the quasi-isothermal halo
  in Equation (\ref{eq:rhoh}) is $\csih = 5$.
    \end{tablenotes}
    \end{threeparttable}
    \label{models}
    \vspace{5mm}
\end{table*}
\renewcommand\arraystretch{1.}

As is well known, the azimuthal velocity field is split in its ordered ($\vphib$) and
dispersion ($\sigmaphi$) components by adopting a generalised Satoh (1980)
$k$-decomposition 
\begin{equation}
      \vphib=k\,\sqrt{\Dels},
      \qquad
      \sigmaphi^2=\sigmas^2+(1-k^2)\Dels,
      \qquad
      \Dels=\vphib^2+\sigmaphi^2 -\sigmas^2;
\label{eq:vphis}
\end{equation}
where $\sigmas$ is the vertical (and then also radial) velocity
dispersion, and the explicit expressions of $\sigmas$ and $\Dels$ are
given in Appendix \ref{app:ABCDEF}.  Therefore, $k=1$ correspond to a
``fast'' rotating galaxy (the isotropic rotator), while $k=0$
describes a galaxy with a flattening totally supported by tangential
velocity dispersion.  In addition to the standard case with constant
$k$, we also explore two more families of rotating galaxies, with a
spatially-dependent Satoh parameter
\begin{equation}
  \ka(r)=k_0+(k_{\infty} - k_0) {s\over \xi_0+s},\qquad
  \ke(r)={\rm e}^{-r/\Rec},
\label{eq:satohkr}  
\end{equation}
with $k_0=0.42$, $k_{\infty}=0.05$, $\xi_0=2.67$. In the exponential
case $\vphib$ decreases significantly at large radii, while the
$\ka (r)$ case $\vphib$ becomes asymptotically flat; in the central
region, instead, stars of a model with $\ke(r)$ rotate rotate almost
as fast as an isotropic rotator, faster than those in the
asymptotically flat case $\ka(r)$, as at the center $\ka(0)=k_0$ is
lower than unity (see Fig. \ref{f1}).

The assumptions of homeoidal expansion, and the neglect of the effects
of the external DM halo on the stellar dynamics inside a few effective
radii of the galaxy (corresponding to more than $99\%$ of the total
stellar mass) was checked by numerical integration of the Jeans
Equations in the full gravitational field, without using the homeoidal
expansion; the integration was done with the multi-component stellar
dynamical code \texttt{JASMINE2} (Jeans AxiSymmetric Models of
Galaxies IN Equilibrium; see Caravita et al. 2021, see also Posacki et
al. 2013).  We found that these effects within $\approx 2\Reff$ are in
fact negligible, so that for the purposes of the present exploration
the formulae in Appendix A can be safely adopted.

To set up realistic galaxy models, we recall that their stellar
central velocity dispersion in absence of the central SMBH, can be
obtained combining Equations (26) and (42) in CMPZ21 (the former with
$\MR=\xi=1$ and $\etag=\etas$, and the latter with $\mu=0$ and
$\etag=0$):
\begin{eqnarray}
  \sigmas^2(0) &=& {G\Mst\MR\over 2\xi\rst}{1-\etas\cos^2\theta\over 1+\etas -2\etas\cos^2\theta}\cr\cr
                        &=&  {G\Mst\MR\over 2\xi (1+\etas)\rst} = {G\Mst\over 2(1-\etas^2)\rst},
\label{eq:sigmazz}
\end{eqnarray}
where the second equality holds when evaluating the
  limit\footnote{The central velocity dispersion of ellipsoidal JJe
    models is discontinuos, with values dependent on the direction
    approaching the center (see for a full discussion CMPZ21).}  along
  the equatorial plane ($\theta=\pi/2$), and finally the last
  expression for minimum halo models, i.e. for $\MR=\MRm$ given in
  Equation (\ref{eq:minhal}). We adopt $\sigmas(0)$ as a proxy for the
  observed velocity dispersion of the galaxy in the central regions
  (outside the sphere of influence of the central SMBH). Moreover,
  from Equation (\ref{eq:phigal}) it follows that the circular
  velocity of JJe models does not vanishes at the center, and
\begin{equation}
      \vgal^2(0)={G\Mst\MR\over \rst\xi}
      =2(1+\etas)\sigmas^2(0), 
\label{eq:vcirc_zero}
\end{equation}
  where the last expression holds independently of the minimum halo model assumption.
  Finally, the model circular velocity in the equatorial plane
  $\vcirc^2(r)=\vbh^2(r)+\vgal^2(r)+\vh^2(r)$ can be written in terms of
  $\vgal(0)$ as
\begin{equation}
{\vcirc^2(r)\over\vgal^2(0)}={\mu\xi\over\MR s}+{\xi\over\xi+s}+
   {\vh^2\over\vgal^2(0)}\left(1-{\csih\over s}\arctan {s\over \csih}\right),  
\label{eq:vcircTOTAL}
\end{equation}
  where we neglect for simplicity the contribution  to the
  gravitational field of the equatorial stellar disk formed by the
  cooling and rotating ISM (see Section 3): notice that in absence of
  the central SMBH,  $\vcirc(0)=\vgal(0)$. If needed, the equation
  above can be recast without difficulty in terms of $\sigmas(0)$, and
  further specialized to the minimum halo case.

\section{The input physics and the hydrodynamical simulations}

The hydrodynamical Equations in the simulations are given in
Equations (1)-(2)-(3) in G19a, where a full discussion of the various
terms is provided. Here we recall the points of direct relevance for
the present paper, and in particular the changes and the additions to
the input physics with respect to G19a.

The {\it mass source terms} for the galactic gas flows are provided by
the mass return from stellar evolution (including mass loss of red giants and AGB stars, SNIa explosions from
the passively evolving population and SNII from the new stars formed,
see Appendix B in G19a, see also Pellegrini 2012, and Ciotti \&
Ostriker 2012), and by cosmological accretion from
a circumgalactic medium (hereafter CGM). Stellar evolution
injects over the galaxy body a total amount of gas of the order of
$\approx 10\%$ of the initial stellar mass, with an almost power-law
steadily declining injection rate $\dot\rho=\alpha(t)\rhost$,
where $\rho$ is the gas density. Instead,
the time-dependence of cosmological mass accretion from the CGM is
modeled following Choi et al. (2017) and Brennan et al. (2018), and
according to Equation (12) of G19a is given by
\begin{equation}
  \dot M_{\rm CGM}(t) = 2\Macc {{\rm e}^{-( t/\tz)^2}\over 1-{\rm
      e}^{-(\Delta t/\tz)^2}}{t\over \tz^2},
\label{eq:Mcgm}
\end{equation}
where we fix $\tz=9$ Gyr, and we scale $\Macc$ so that the total
mass accreted from the CGM is $\simeq 0.44\Mst$ during the time span
on the simulation, $\Delta t = 12$ Gyr.

The various source terms are injected into the galaxy, not only mass, but also
{\it momentum}, {\it internal}, and {\it kinetic energy}; the
associated terms are given in Equations (52)-(53) in G19a (Negri et
al. 2014a,b; Ciotti et al. 2017, see also Chapter 10 in Ciotti
2021). In particular, the dynamical properties of the stellar
component enter in the thermalization term in the energy equation as
\begin{equation}
\dot E_{S}=\dot\rho\, {{\rm Tr}\,\sigma^2 +\Vert{\bf u} -\vphib {\bf
    e}_{\varphi}\Vert^2\over 2}
\label{eq:Esource}
\end{equation}
where ${\rm Tr}\,\sigmas^2=2\sigmas^2+\sigmaphi^2
=3\sigmas^2 + (1-k^2)\Deltas$ is the trace of the velocity dispersion
tensor, 
${\bf u}$ is the fluid velocity, and $\vphib$ the azimuthal
streaming velocity of stars in Equation (\ref{eq:vphis}). Similarly, it
can be proved that the momentum source term is given by
\begin{equation}
  \dot {\bf m}_{S}=\dot\rho\,\vphib {\bf e}_{\varphi}.
  \label{eq:momsource}
\end{equation}
Also the mass accretion flow from the CGM imposed at the outer
boundary of the numerical grid injects energy and momentum in the
computational domain. We assume a purely radial accretion velocity at the outer
grid boundary (at $\rt=250$ kpc), so no angular momentum is associated with 
$\dot{M}_{\rm CGM}$, and the modulus of this infall velocity is
\begin{equation}
v_{\rm CGM} = \sqrt{-{\phigal(\rt)-\vh^2(\rt)\over 2}}.
\end{equation}
This value corresponds to half of the free-fall velocity from
infinity, under the assumption that the DM quasi-isothermal halo in
Equation (\ref{eq:rhoh}) is truncated at $\rt$.  Besides the mass
input rate and infall velocity, the numerical modeling also requires
the angular distribution and the temperature of the infalling
material. Following G19a, its internal energy is set so that its sound
velocity equals $v_{\rm CGM}$, while the CGM mass flux is weighted by
a $\sin^2\theta$ angular dependence, therefore most of the CGM is
injected near the equatorial plane. Finally, the metallicity of the
CGM is obtained by assuming $\dot{M}_{\rm CGM}$ made of $1/4$
primordial gas, and $3/4$ low metallicity gas of $0.2$ solar abundance
(see also Table 1 in G19b).

\begin{figure}
\hskip -0.5truecm 
\includegraphics[width=0.38\linewidth, keepaspectratio]{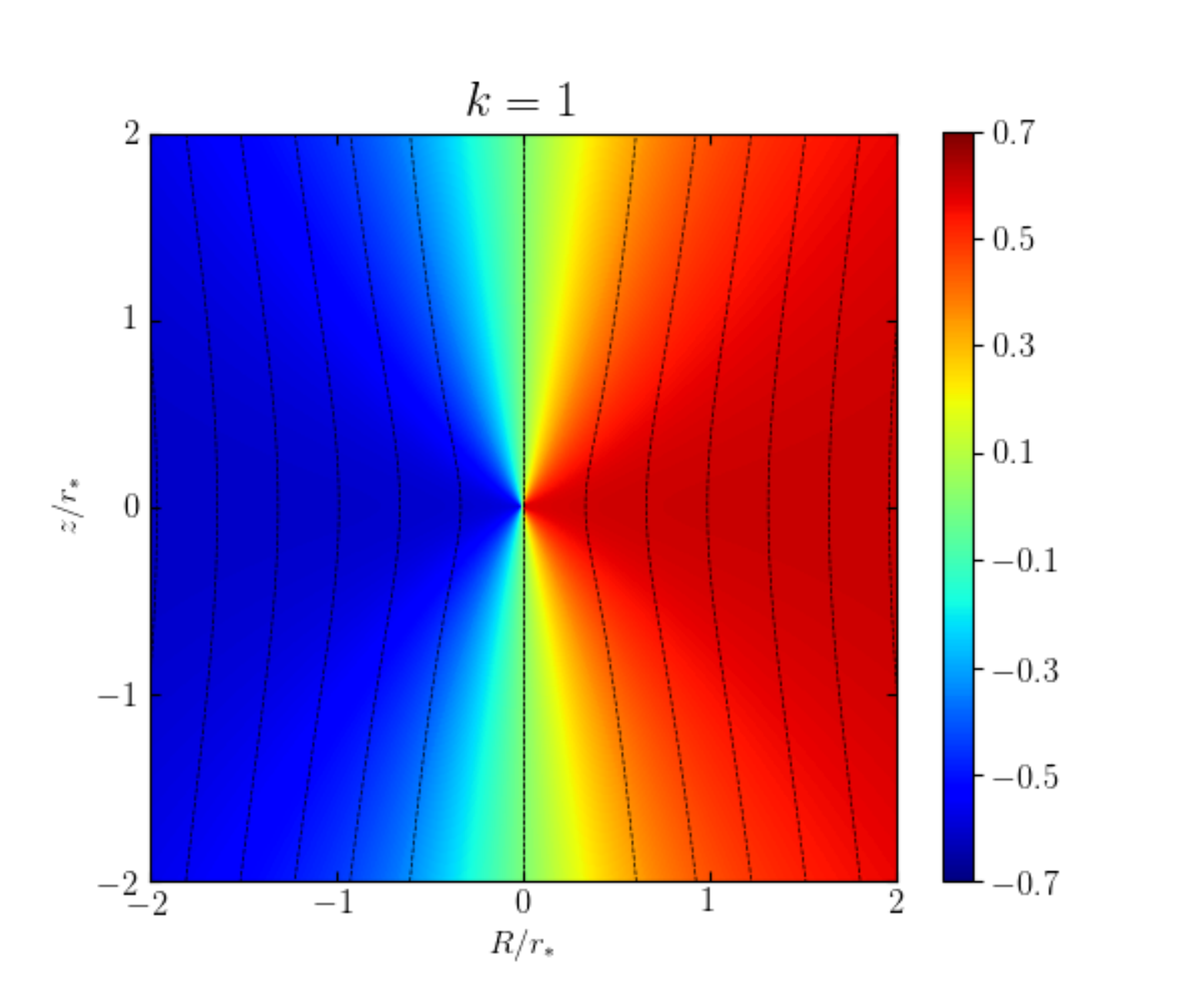}  
\hskip -0.7truecm 
\includegraphics[width=0.38\linewidth, keepaspectratio]{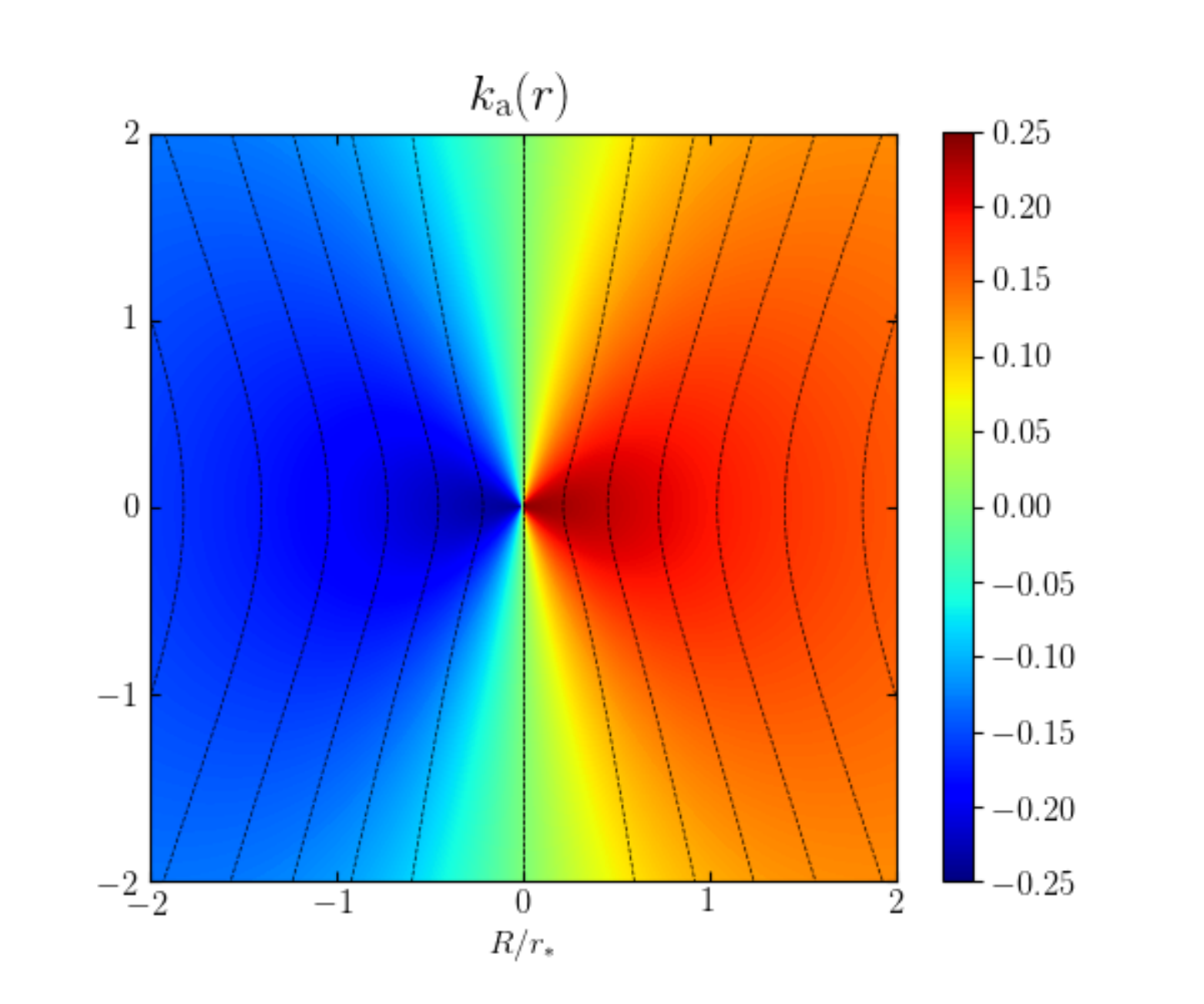}
\hskip -0.7truecm 
\includegraphics[width=0.38\linewidth, keepaspectratio]{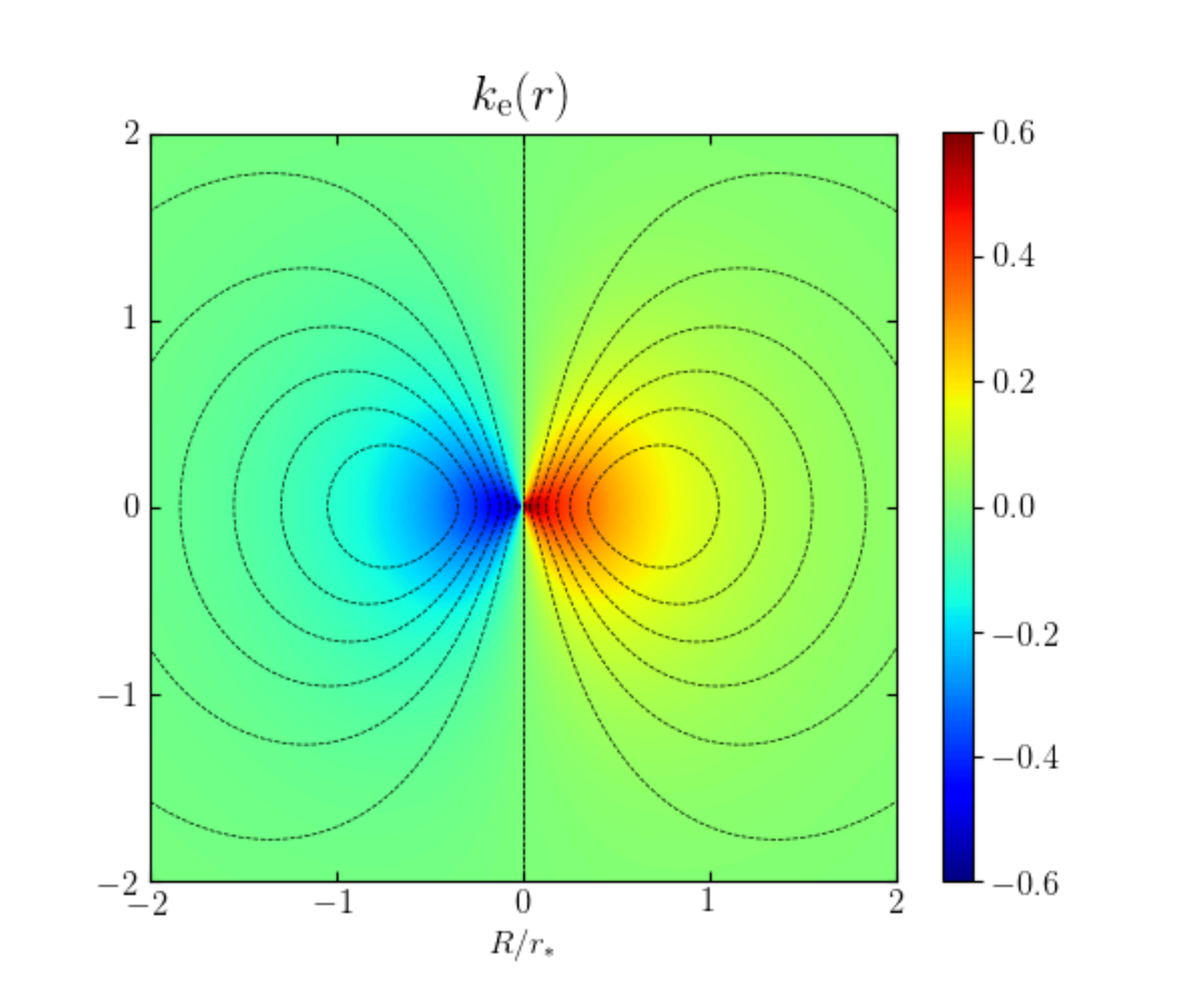}
\caption{Maps of the stellar ordered rotational velocity field
  $\vphib/\sqrt{\phin}$ in the $(R,z)$ plane, for the isotropic
  rotator (left), and for the two spatially dependent Satoh
  decompositions in Equation (\ref{eq:satohkr}); $\phin=G\Mst/\rst$ is
  given in Table~\ref{models} for the different model families.  The
  dotted lines are contours of constant angular momentum per unit mass
  of the stellar component; as shown in Equation (\ref{eq:djzdt}), in
  absence of mass sources and viscous dissipation, or for a gaseous
  halo rotating with the same velocity $\vphib$ of the stars, the
  cooling gas would fall at $\Rin$ on the equatorial plane along these
  lines, and then contract to $\Rfin$ as illustrated in the right
  panel of Figure \ref{f1}.}
\label{f2}
\end{figure}
\vskip 0.8truecm 

In rotating models, we follow the evolution of the equatorial cold
gaseous disk produced by the gas inflow and cooling, modeling the star
formation in it, and the consequent gas accretion on the SMBH
associated with the (local) Toomre instability. From Equations
(13)-(14) in G19a, we evaluate at each time-step the $Q$-profile of
the disk as:
\begin{equation}
  Q(R)={c_s\kappa\over \pi G\Sigma},\qquad
  \kappa^2 = {2\Omega\over R}{d (\Omega R^2)\over dR},\qquad
  \Omega(R)={\vcirc(R)\over R},
\end{equation}
where $\Sigma$ is the gas surface density of the disk, $c_s$ is the
sound velocity, and $\vcirc$ is the circular velocity in the
equatorial plane given by Equation (\ref{eq:vcircTOTAL}).  When the
Toomre instability affects (a ring) in the cold gaseous disk, we
assume that a fraction $\Delta Q = \max(1-Q,0)$ of the unstable gas
falls onto the center, on a time scale given by the local $\vcirc(R)$
as in Equation (15) in G19a, which will result in decrement of
$\Sigma$ (thus increment of $Q$). We refer to G19a also for a
description of the algorithm for the numerical treatment of
instability, and the associated redistribution of mass, energy, and
angular momentum, as well as of the disk $\alpha$-viscosity.  In this
way, $Q$ is re-established to unity, and the disk self-regulates
locally (Bertin and Lodato 1999, Cossins et al. 2009).

Disk instability induces also star formation, and according Equation
(20) in G19a,
\begin{equation}
  \dot\rho_{*,Q}= \eta_{\rm SF,Q}\,\Delta Q\,\rho\,\Omega,\qquad
  \Delta Q=\max(1-Q,0),\qquad
  \eta_{\rm SF,Q}=0.02,
\end{equation}
where we reduced $\eta_{\rm SF}$ by a factor of 5 respect to the value
of $0.1$ adopted in G19a. The IMF of star formed in the disk is
assumed to be top-heavy (e.g., see Goodman and Tan 2004) to match the
IMF seen in the central disk of MW and M31, and we assume an initial
mass function for stars of mass $M$, formed in the unit time, at time
$t$, of the form:
\begin{equation}
  {d N\over dM}={N_0 (t)\over\Msol}\left({M\over\Msol}\right)^{-1.65}, 
\end{equation}
with $\Msol <M<50\Msol$, and $N_0(t)$ determined to match the total 
mass of disk stars formed in the time step. Such an IMF gives 
$\approx 60\%$ of the total new star mass in massive stars 
($M > 8\Msol$), which will turn into SNe II on a timescale of 
$\approx 2\times 10^7$ yrs. 

Disk instability is not the only channel considered for star
formation. In the simulations we also allow for star formation
provided that 1) the gas temperature falls below $4\times 10^4$ K,
{\it and} 2) the gas density is higher than $10^5$
atom/cm$^{-3}$. When the temperature and the density of a gas element
satisfy the conditions above, star formation takes place via Jeans
instability with the standard timescale given by
$\max(\tau_{\rm cool},\tau_{\rm dyn})$, as fully described in Equations
(22)-(23) in G19a.

A new feature of the present simulations is the gravitational effect
on the gas flows due to the stellar disk of new stars formed by the
rotating cooling gas. In fact, albeit the total mass of the disk at
any time is much lower than the total initial galaxy mass (stars plus
DM), its gravitational field can be important in the central galactic
region, especially near the equatorial plane.  Two competitive effects
of the compression produced by the vertical gravitational field of the
stellar disk on the accreting gas are expected: one is compressional
heating, with a reduction of accretion, the other is the tendency
towards gas cooling and accretion, due to the increase in gas
density. In past simulations of rotating gas flows (e.g., Negri et
al. 2014a,b; Ciotti et al. 2017), only the second effect could be at
work, as the gravitational field of the stellar disk was not taken
into account. We consider here a semi-quantitative modelization of the
disk that allows for a fast numerical computation. In practice, at
each time-step, we compute the time dependent disk stellar mass
$\Mdstar (t)$, and the half mass disk radius $\Rds (t)$ from the
history of star formation (see Table \ref{results}), then we assume
that the disk is described by a Kuzmin-Toomre razor thin disk (see
Binney and Tremaine 2008)
\begin{equation}
  \Sigds (R)={\Mdstar a\over 2\pi (R^2+a^2)^{3/2}},\quad
  \phids(R,z)=-{G\Mdstar\over\sqrt{R^2+(a+\vert z\vert)^2}},
\label{eq:phidisk}
\end{equation}
where $a=\Rds/\sqrt{3}$. The formula above is used to compute and
update at each time-step the vertical and radial gravitational fields
produced by the stellar disk. For simplicity we do not compute the
gravitational field due to the gaseous equatorial disk\footnote{From
  Table \ref{results} notice how the surface density of the gaseous
  disk is significantly lower than the surface density of the more
  concentrated stellar disk, so that its vertical gravitational field
  is correspondingly weaker.}, nor the modifications of the stellar
kinematics produced by the (time-dependent) gravitational field of the
stellar disk; instead we take into account the
change in the total gravitational field due to the growth of the
central SMBH and the decrease of stellar mass (see Appendix B for more
details).

Here we list the main additions/changes adopted in the present
simulations. Following the treatment in N\'u\~nez et al. (2017), we
now also consider the effect of UV heating produced by the (massive)
new stars formed in the disk, updating the ISM temperature as
\begin{equation}
{dT\over dt}={10^4\,{\rm K}-T\over t_{\rm rec}},\quad {\rm if}\quad T\leq
10^4\,{\rm K}
\end{equation}
where the recombination time-scale $t_{\rm rec}$ is estimated as 
\begin{equation}
t_{\rm rec}\equiv {1\over n_{\rm H}\alpha_{\rm B}},\quad \alpha_{\rm
  B}\simeq 2.56\,10^{-13}\;{\rm cm}^{-3} {\rm s}^{-1},
\end{equation}
where $n_{\rm H}$ is the hydrogen number density of the ISM in
cm$^{-3}$, and $\alpha_{\rm B}$ is the effective radiative
recombination rate for hydrogen, assuming a gas temperature of $10^4$
K (Draine 2011). The UV heating is effective in each grid, provided that 1)
the temperature is less than $10^4$ K, {\it and} 2) and the numerical
grid size is smaller than the Stromgren sphere, estimated from
Equation (3) of N\'u\~nez et al. (2017).

A key ingredient of the hydrodynamical simulations is represented by
the input physics describing energy and momentum feedback from the
stellar components, and from accretion events on the central MBH.  For
a complete description of the input physics and its numerical
implementation we refer to Section 2.7 in G19a, and Appendix A and B
therein. (1) we adopted a maximum wind efficiency of $\epswM=0.005$ as
in G19a\footnote{Notice that in G19b,c $\epswM= 0.0015$.}, (2) we
increased the opening angle of the AGN winds by weighting its angular
distribution by $\vert\cos\theta\vert$ (rather than $\cos^2\theta$ as
in G19a); and (3) we smoothed the transition from the cold to the hot
AGN feedback mode by introducing two correction factors ($A=0.5$,
$B=0.5$) to Equations (27) and (28) in G19a as follows:
\begin{equation}
\epsw =  \epswM \sqrt{{5\,l\over 4+l}\,
 {\rm e}^{-\left(A\,{\dot{M}_{\rm disk,crit}}/{\dot{M}_{\rm BH}}\right)^4}}
\end{equation}
\begin{equation}
  \dot{M}_{\rm w}  = \dot{M}_{\rm disk}\times\left( 1- B\,\sqrt{{3r_s\over r_{\rm tr}}}\right).
\end{equation}

As a check, we performed several numerical experiments, at different
spatial resolutions (up to a factor of $10$ higher), and with
different choices for the parameters modeling AGN feedback/star
formation/CGM accretion. In general these changes produce results in
the expected direction, and overall the presented models, although can
surely be improved in some specific aspect, are well representative of
the results that can be obtained in the present framework.

\subsection{The numerical code}

We solve the Eulerian hydrodynamical equations, together with those
relative to 12 metal tracers (G19b) and grain physics (G20), with our
high-resolution {\it MACER} (Massive AGN Controlled Ellipticals
Resolved) grid hydrodynamical code (G19a), based on the Athena++ code
(version 1.0.0; Stone et al. 2020). We use spherical coordinates
$(r,\theta)$ and we assume axi-symmetry, but allow for rotation
(a.k.a. 2.5-dimensional simulation). The outer boundary is chosen as
$250$ kpc from the galaxy center to well enclose the whole stellar
distribution of the galaxy, and also a significant region of the
group/cluster DM halo. The inner radial grid point $r_{\rm in}$ is
placed at $25$ pc from the galaxy center, allowing us to resolve the
fiducial Bondi radius; for example, the Bondi radius of the three
families of models in Table \ref{models}, evaluated for a reference
gas temperature of $T=10^6$ K, and an initial SMBH mass of
$\Mbh=0.001\Mst$, is $\simeq 30$ pc, $\simeq 65$ pc, and $\simeq 150$
pc, respectively for LM, MM, and HM models. Of course, as the SMBH
mass increases with time, the numerical resolution tends to improve as
the simulations proceed. Even if this resolution is quite high when
compared to that adopted in other numerical studies, for some tests
(see below) we also performed significantly more time-expensive
simulations, with $r_{in}=2.5$ pc. The radial grid is logarithmic,
with 120 grid points and an expansion factor of
$\Delta r_{i+1}/\Delta r_i = 1.1$ between two adiacent grids. The
azimuthal angle $\theta$ is divided into 30 uniform cells, and covers
an azimuthal range from $0.05\pi$ to $0.95\pi$. The numerical solver
for the gas dynamics is composed by the combination of the HLLE
Riemann Solver, the PLM reconstruction, and the second-order van Leer
integrator.  Outflow boundary conditions are imposed at the galaxy
outskirts, which allows the gas to escape from the galaxy, but does
not force it to do so.  The inner boundary conditions are designed to
allow for the ISM to flow inward freely, and to avoid mass outflow
from the center, while the treatment of the AGN winds is implemented
at the innermost {\it active} cells, placed immediately outside the
inner boundary radius.

\section{Exploring the parameter space: results}

>From the description of the models in Section 2 and of the input
physics in Section 3, it should be clear that a systematic and
complete exploration of the parameter space is impossible. In fact, a
run of a model with the inner grid placed at 25 pc from the origin
takes around 3-4 days with 40 cores ($2\times$ Skylake 6148 on a
single node), while it takes $10\times$ longer time with the increased
resolution and the first grid placed at 2.5 pc from the center. For
this reason we fixed the galaxy flattening to represent E3 galaxies,
and we consider three representative values for the initial stellar
mass, i.e.  $\Mst = 1.5\times 10^{11}\Msol$, $3.4\times 10^{11}\Msol$,
and $7.8\times 10^{11}\Msol$; the explored models (respectively LM,
MM, and HM in Table \ref{models}) correspond to galaxies that are
massive enough that the evolution of the gaseous halo is not entirely
dominated by SNIa heating (e.g., Ciotti et al. 1991), being smaller
systems able to sustain galactic winds just due to the SN energy
input. The models are constructed to be on the Fundamental Plane of
elliptical galaxies, and as in our previous works the age of the
galaxy at the beginning of the simulation is fixed to be 2 Gyr, so
that the initial phases of galaxy formation are terminated (and a SMBH
with a mass near to observed values is assumed to be be already in
place). The galaxy DM halo corresponds to minimum halo models, with a
mass 18 times larger than the initial stellar mass, and a scale length
$\simeq 13$ times larger than that of the stellar distribution; in
this way, the galactic DM halo is very well represented by a NFW-like
profile over a very large radial range, down to the galaxy center. The
group/cluster DM halo is instead important only at very large radii
(outside several effective radii of the galaxy), with asymptotic
circular velocity fixed to match the circular velocity near the center
(in absence of central SMBH).  All the structural parameters of the
models are given in Table \ref{models}. Finally, as detailed in Table
\ref{results}, for each of the three mass models, we consider three
different rotational supports: no rotation (all the galaxy flattening
is due to tangential velocity dispersion), moderate rotation (rotation
exponentially declining in the outer regions as described by Equation
(\ref{eq:satohkr})), and finally the isotropic rotator case (all the
galaxy flattening is supported by ordered rotation).

\renewcommand\arraystretch{1.4}
\begin{table*}
\centering 
\caption{Integrated properties of the models at 13.7 Gyr}
\vspace{2mm}
\begin{threeparttable}
\begin{tabularx}{\textwidth}{c@{\extracolsep{\fill}}ccccccccccccccc}
  \toprule
  Model name
  &  $\Delta\Mbh$
  &  $\MdHI$
  &  $\RdHI$
  &  $\SdHI$  
  &  $\Mdstar$
  &  $\Rds$
  &  $\Delta\Mst$
  &  $\Mout$
  &  $\Mhot$
  &  $\Lx$
  &  $\Tx$\\
  & $(10^{8}\Msol)$
  & $(10^{8}\Msol)$
  & $(\kpc)$
  & $(\Msol/{\rm pc}^2)$
  & $(10^{8}\Msol)$
  & $(\kpc)$
  & $(10^{8}\Msol)$
      & $(10^{8}\Msol)$
   & $(10^{8}\Msol)$
   & $(10^{40}{\rm erg}/{\rm s})$
  & $(10^6\, {\rm K})$\\
     & (1)  & (2)    & (3)   & (4)      & (5) & (6)   & (7)  & (8) & (9)  & (10)   & (11) \\  
   \midrule                                                 
  LM$_0$ & 7.0  & 0.0      & 0.0    & -          &0.0& 0.0   & 0.0    &204.5 &5.6 & 5.4   & 6.1 \\  
  LM$_k$ & 12.8& 2.4      & 0.7    &150.3    &2.1&0.1    & 4.8    &507.1 & 4.3 & 0.8 & 6.9\\
  LM$_1$ & 16.7 & 60.3    & 4.4  &101.4     &3.0&0.3    & 6.8    &523.0 &1.6 & 0.1 & 11.0 \\
  \midrule 
  MM$_0$ & 22.2 & 0.0       & 0.0   &-        &0.0& 0.0   & 0.0     &156.6& 49.9 & 20.3 & 10.9 \\
  MM$_k$ & 36.7 & 11.0   & 0.5   &1454    &5.3& 0.1   & 12.2   &1181.5&21.1 & 8.6 & 9.4 \\
  MM$_1$ & 71.9 & 46.6   & 3.6   &114.4   &12.2&0.3  & 28.0  &1236.8&11.6 & 1.1 & 11.5\\
  \midrule 
  HM$_0$ & 85.1 & 0.0       & 0.0   &-          &0.0  & 0.0 & 0.0   & 3273.8  &76.4& 12.9 & 12.5 \\
  HM$_k$ & 90.2 & 12.3   & 0.6   &1167    &12.5 & 0.1& 28.9   &2186.1   &240.1& 87.3 & 12.3 \\
  HM$_1$ & 143.0& 57.8 & 3.0    &208.6   &29.8 & 0.3& 68.7   &2833.6    &117.9 &18.3 & 12.8\\
  \bottomrule
\end{tabularx}
\vspace{1.5mm}
\begin{tablenotes}[para,flushleft]
  \footnotesize Final values of a selection of global properties for
    the models on the leftmost column; the subscript in the model name
    indicates the adopted parameterization azimuthal stellar motions,
    in order of increasing importance of the rotational support: $0$
    means no ordered rotation, $k$ indicates the exponentially
    declining ordered rotation as given by $\ke(r)$ in Equation
    (\ref{eq:satohkr})), and $1$ the isotropic rotator. The other
    columns give: (1) the accreted SMBH mass, 2) the cold
    ($T\leq \Tc= 5\times 10^5$ K) gas mass in the equatorial gaseous
    disk, (3) the cold disk truncation radius, (4) the cold disk
    average surface density, (5) the stellar mass of the equatorial
    disk, (6) the half-mass radius of the stellar disk, (7) the total
    mass of star formed in the galaxy, (8) the total gas mass ejected
    from the numerical grid (250 kpc), (9) the total mass of the hot
    ISM (defined as the gas with $T > \Tc$ and $r < 5\,\Rec$), (10)
    the X-ray luminosity $\Lx$ of the ISM (in the $0.3-8$ keV energy
    band, in the region bounded by $100\, {\rm pc} < r< 5\,\Rec$), and
    (11) the $0.3-8$ keV emission-weighted temperature $\Tx$ in the
    same region.
\end{tablenotes}
    \end{threeparttable}
\label{results}
\end{table*}
\renewcommand\arraystretch{1.}
\vspace{5truemm}

\subsection{SMBH accretion and duty-cycles}

\begin{figure*}
  \includegraphics[width=0.48\linewidth, keepaspectratio]{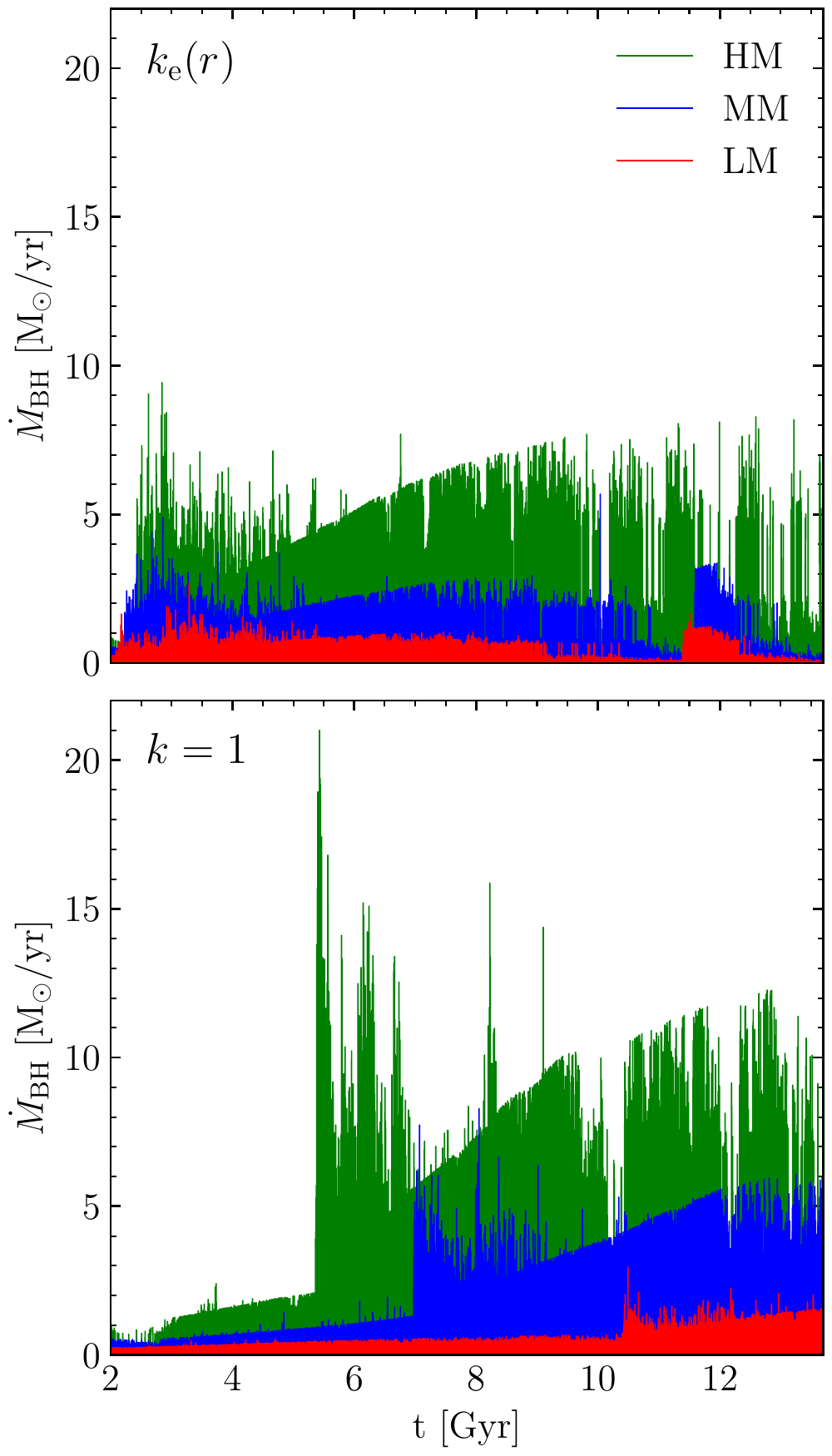}
  \includegraphics[width=0.48\linewidth, keepaspectratio]{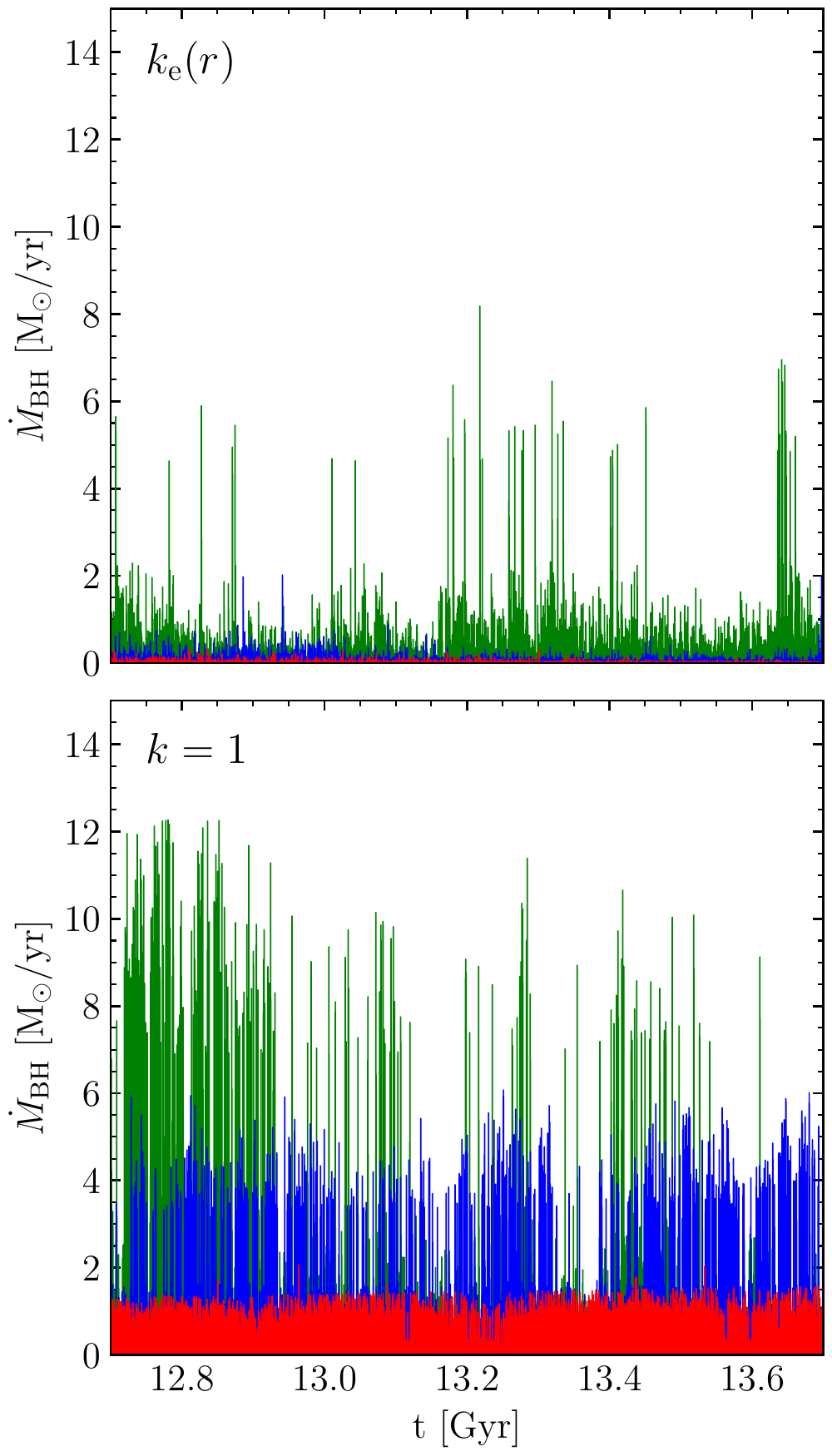}  
  \caption{Time evolution of the SMBHs accretion rate, for the
    high-mass (green), medium-mass (blue), and low-mass (red) galaxy
    models, during the whole evolution (left panels), and over the
    last Gyr (right panels). $\dMbh$ spans a range of
    $\approx (10^{-5}-10^{-1})\times\dMedd$, with very few
    accretion episodes with $\dMbh$ exceeding
    $\dMedd$. The different stellar ordered rotation is
    indicated by the $k$ values in the upper left corner (with
    $\ke(r)$ corresponding to mild rotation, and $k=1$ to
    isotropic rotators). }
\label{f3}
\end{figure*}

>From inspection of Table \ref{results}, we found systematic trends
between the mass $\Delta\Mbh$ accreted by the SMBH at the end of the
simulations, and the galaxy mass and the degree of internal ordered
rotation.

The first trend  is that $\Delta\Mbh$ increases with galaxy mass. This is not
surprsing, as the mass losses from stars scale linearly with the
stellar galaxy mass $\Mst$, and from Equation (\ref{eq:Mcgm}) also the
mass accretion from the group/cluster environment scales linearly with
the galaxy mass, so that in more massive galaxies more gas is
available for accretion. However, from inspection of the $\Mst$ values in Table
\ref{models}, one sees that $\Delta\Mbh$ increases more than
linearly with the mass sources, i.e., SMBHs in massive galaxies
accrete more efficiently than SMBHs in galaxies of lower mass. This is
a quite well established result, a natural byproduct of the larger
binding energy per unit mass of more massive galaxies, as dictated by
the Faber-Jackson law, which leads to a more efficient gas retention,
as the heating sources (thermalization of stellar winds, and SN
explosions) scale instead linearly with the galaxy mass (e.g., Ciotti
et al. 1991). This is confirmed by the amounts of hot
gas retained by the galaxies inside a volume of $5\,\Rec$ at the end
of the simulations (see Column 9 in Table \ref{results}, see also
Figure \ref{f7}).

\begin{figure*}
  \includegraphics[width=0.48\linewidth, keepaspectratio]{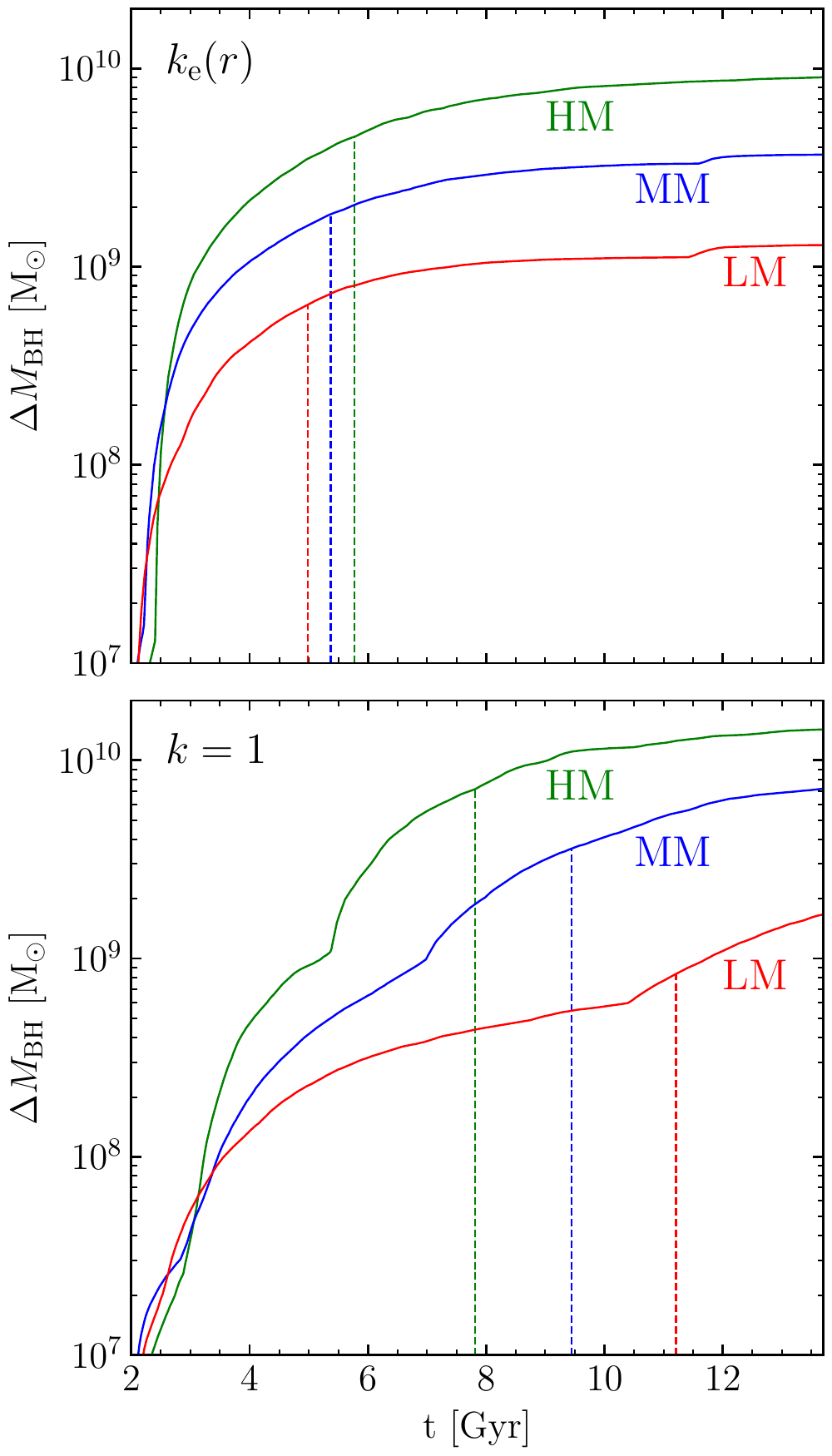}
  \includegraphics[width=0.48\linewidth, keepaspectratio]{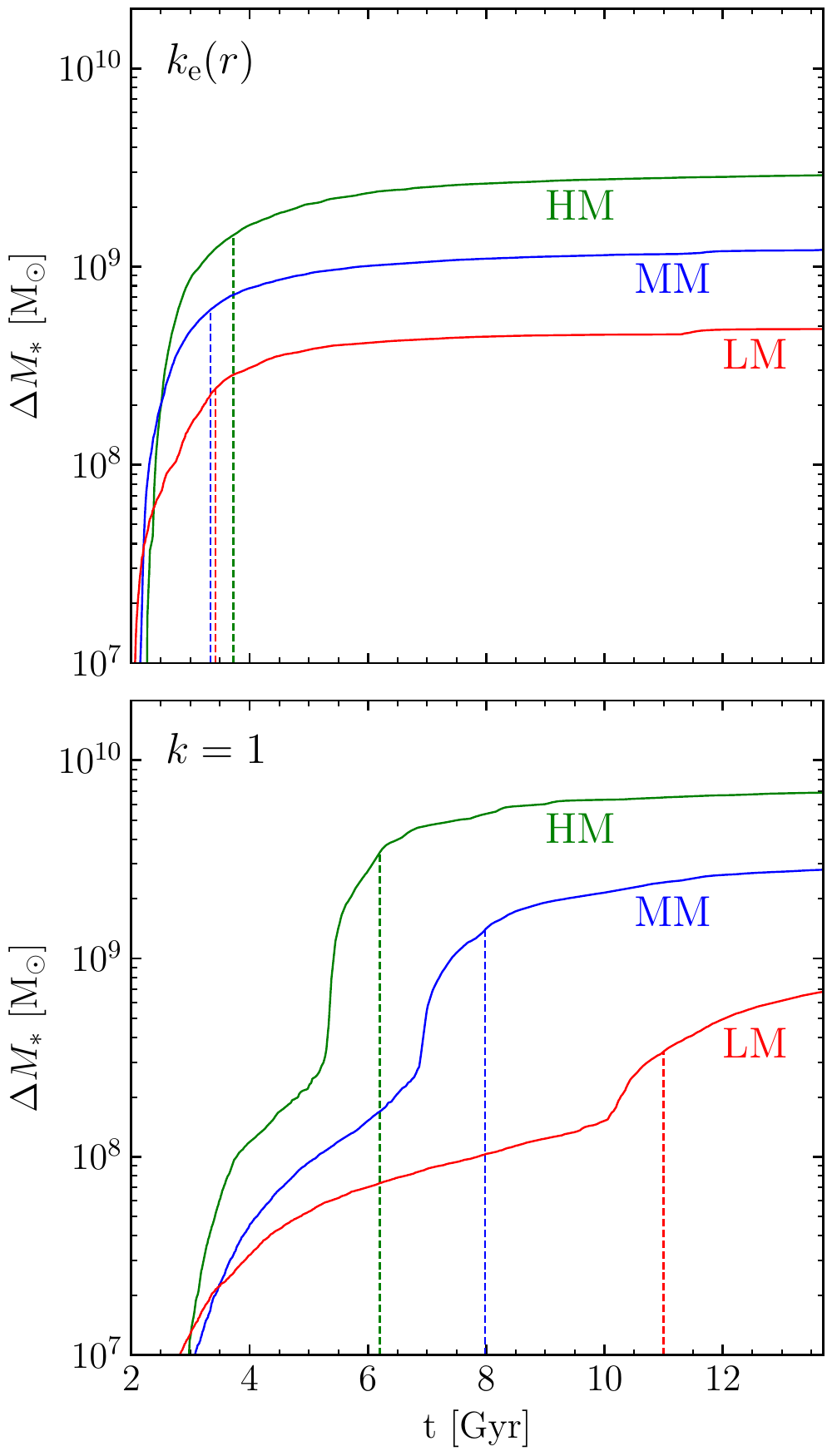}  
  \caption{Left panels: time evolution of the mass $\Delta\Mbh$
    accreted by the central SMBH, for the same models in Figure
    \ref{f3} (high-mass: green, medium-mass: blue, low-mass:
    red). Right panels: evolution of the time-integrated star
    formation rate, $\Delta\Mst$, for the same models in Figure
    \ref{f6}; notice that $\Delta\Mst$ is not the present-day mass of
    stars formed during the model evolution (cfr. Columns 5 and 7 in
    Table \ref{results}), as a significant fraction of $\Delta\Mst$ is
    re-injected in the ISM from mass losses from the newly formed
    stars. In each plot, the vertical lines mark the time at which
    each quantity reaches half of its final value.
    Note haw BH accretion occurs later in lower mass galaxies.}
\label{f4}
\end{figure*}

The second trend is that, in each of the families (LM, MM, and HM),
the more rapidly rotating galaxies accrete more material on to their
SMBH (see the $\Delta\Mbh$ evolution in Figure \ref{f4}, left
panels). This result may appear at odds with expectations, as the
centrifugal barrier of faster rotating galaxies acts in the sense of
preventing accretion (see Figure \ref{f2}). In fact quite the opposite
happens: a stronger rotational favour large scale instabilities and
gas cooling over the galaxy body, leading to stronger inflows on the
equatorial plane, and to the formation of more massive and extended
gaseous disks than in mildy rotating models, where less massive and
smaller disks form (see Columns 2 and 3 in Table \ref{results}, see
also Section 4.2).  Toomre instabilities then discharge gas on to the
central SMBH, following the prescriptions of Section 3; interestingly,
the smaller disks have a higher gas density (Column 4 in Table
\ref{results}), and are thus more prone to Toomre instability than the
more massive and more diffuse gaseous disks of faster rotating
models. A check shows that the larger $\Delta\Mbh$ of fast rotators is
due to fewer instability events, characterized though by significantly
larger mass accretion episodes.

In Figure \ref{f3} we show the time evolution of $\dMbh$ over the
whole time interval spanned by the simulations (left panels), and over
the last Gyr (right panels). In the top panels the plots refer to the
midly rotating models, while in the bottom panels to the isotropic
rotators. The dependence of the SMBH accretion rate on galaxy mass and
internal rotation is clearly detectable: the accretion episodes reach
systematically higher $\dMbh$ in high mass models and in models with
substantial internal rotation. The left panels also show how important
accretion episodes begin almost immediately in the mildy rotating
galaxies (top panel), while the first massive accretion episodes in
the isotropic rotators (with peaks of
$\dMbh\simeq 10-20\; \Msol {\rm yr}^{-1}$) start at quite late times,
with the epoch of the first important event increasing at decreasing
galaxy mass (bottom panel), with more rotating gas collects at larger
radii and lower densities, hence lower lower cooling and later
accretion events.  At low redshift, peak rates of accretion hardly
reach Eddington values, $\dMedd =L_{\rm Edd}/0.1c^2$, with
common values of $\dMbh$ in the range
$\approx (10^{-5} - 10^{-1})\times \dMedd$.

In the left panels of Figure \ref{f4} we plot the function
$\Delta\Mbh=\int\dMbh dt$ as a function of time, where the
vertical lines mark the time at which half of the final value of
$\Delta\Mbh$ is reached. The more conspicuous features are the more
rapid growth in the mildly rotating models (top panel) than in the
isotropic rotators (bottom panel); the jumps of $\Delta\Mbh$ in the
isotropic rotators (corresponding to the jumps in $\dMbh$ in
Figure \ref{f3}, bottom left), absent in the less rapidly rotating
galaxies;, and finally the {\it inversion} of the time order in which
half of the accreted mass is reached, with a faster evolution of the
HM model with respect to the LM one, in the isotropic rotator case,
while the opposite holds for the $k_{\rm e}(r)$ models. The bottom
panels of Figure \ref{f4} are consistent with the observed fact that
lower mass Seyfert galaxies peak at later epochs than do higher mass
Quasars, a dramatic confirmation of our modeling.

Overall, the results in this Section confirm that AGN feedback is
efficient to maintain SMBHs masses in the present universe small, when
compared to the available gas that could be accreted with unstopped
cooling flows (approximately two orders of magnitude more than the
final SMBHs masses, even not considering group/cluster accretion). It
is also shown how specific properties of ordered rotation can
significantly affect the accretion history and the AGN feedback in
ETGs. Finally, we notice that the final SMBHs masses obtained in the
present simulations are somewhat larger than the observed
ones. However, our test models run at higher resolution (with the
first radial grid point placed at 2.5 pc from the SMBH, instead of 25
pc as in the model survey here presented) indicate that the final
$\Delta\Mbh$ mass would be appreciably smaller in a still higher
resolution simulation, with a significant fraction of the mass that in
the quoted simulations falls to the SMBH instead being either ejected
or turned into stars. Thus the too large final SMBH masses in the
present simulations would probably be reduced to values consistent
with the Kormendy and Ho (2013) relation, were we able to proceed to
still higher resolution simulations.

\subsection{The equatorial gaseous and stellar disks. Star formation
  rates}

\subsubsection{The equatorial disks}

With the exception of non rotating models, all models in Table
\ref{results} are characterized by different degrees of internal
ordered rotation (Section 2.1).  It is a natural result of gas cooling
in the presence of angular momentum that even in case of
low-rotational support of the stellar component, cold gaseous disks
form in the equatorial plane of the galaxy. This is because mass
injection from the stellar population contributes a source of momentum
and angular momentum for the ISM proportional to the local streaming
velocity of stars $\vphib$ (e.g., see Equation (53) in G19a, Chapter
10 in Ciotti 2021b).  Several works have explored the problem of
rotating cooling flows, both numerically with the aid of
hydrodynamical simulations (e.g., see Brighenti and Mathews 1996,
D'Ercole \& Ciotti 1998, Negri et al.~2014a,b, Negri et al.~2015), and
analytically (Ciotti \& Pellegrini 1996, Posacki et al.~2013).  In the
above investigations, no AGN feedback was considered. The common
findings can be summarized as (1) a substantial and enhanced tendency
of the rotating ISM towards instabilitites/cooling (almost absent in
non rotating models), (2) a rotational field of the ISM comparable to
that of the stars, with the ISM rotational velocity
$u_{\varphi}\simeq\vphib$, (3) the formation of cold gaseous disks in
the equatorial plane, more or less massive and extended depending on
the amount of ordered rotational support, (4) a substantial decrease
of the ISM X-ray luminosity $\Lx$ when compared to that of similar
galaxies in absence of rotation. This latter result is interesting, as
observations (e.g., see Sarzi et al. 2013, Juranova et al. 2020) seem
in fact to indicate that rotating systems tend to be X-ray
underluminous when compared with non rotating galaxies of similar
optical luminosity.

In the previous studies two major ingredients were missing, namely the
effect of disk instabilities/viscosity, and AGN feedback. The two
phenomena are clearly related, as in a rotating system the centrifugal
barrier would make accretion on the SMBH impossible in the absence
viscous effects.  We studied in exploratory works the combined effect
of rotation and AGN feedback (Ciotti et al.~2017, Pellegrini et
al.~2018, Yoon et al.~2018, G19a,b, G20), with a phenomenological
description of Toomre instability, angular momentum migration, and
mass discharge on the SMBH. In the present study we adopt more
realistic galaxy models, an updated treatment of disk instabilites and
gas viscosity, and an improved AGN feedback modelization.  Overall,
for the comprehensive set of rotating models in Table \ref{results}
the four main results mentioned above are recovered.

\begin{figure}
  \includegraphics[width=0.48\linewidth, keepaspectratio]{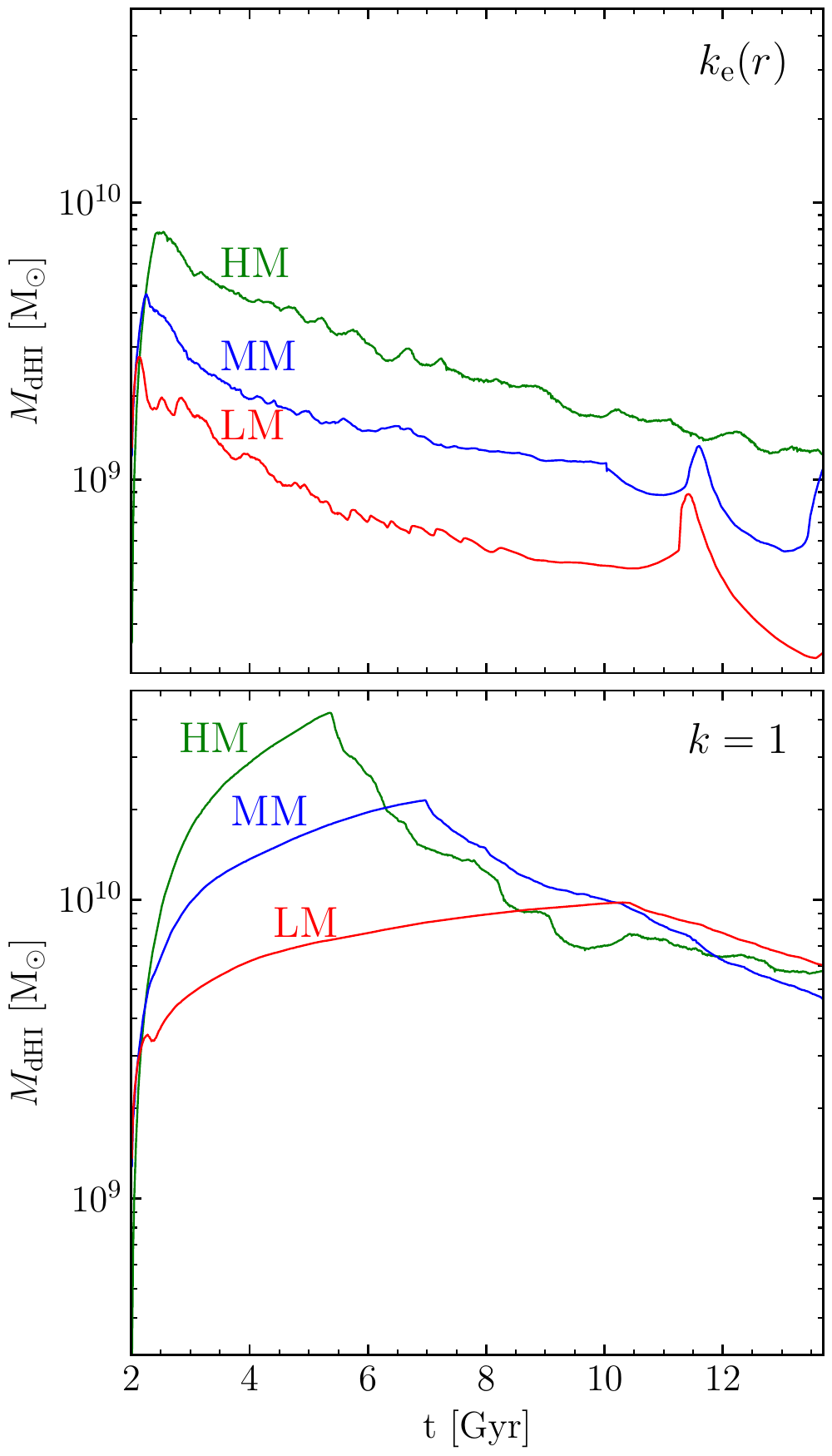}
  \includegraphics[width=0.48\linewidth, keepaspectratio]{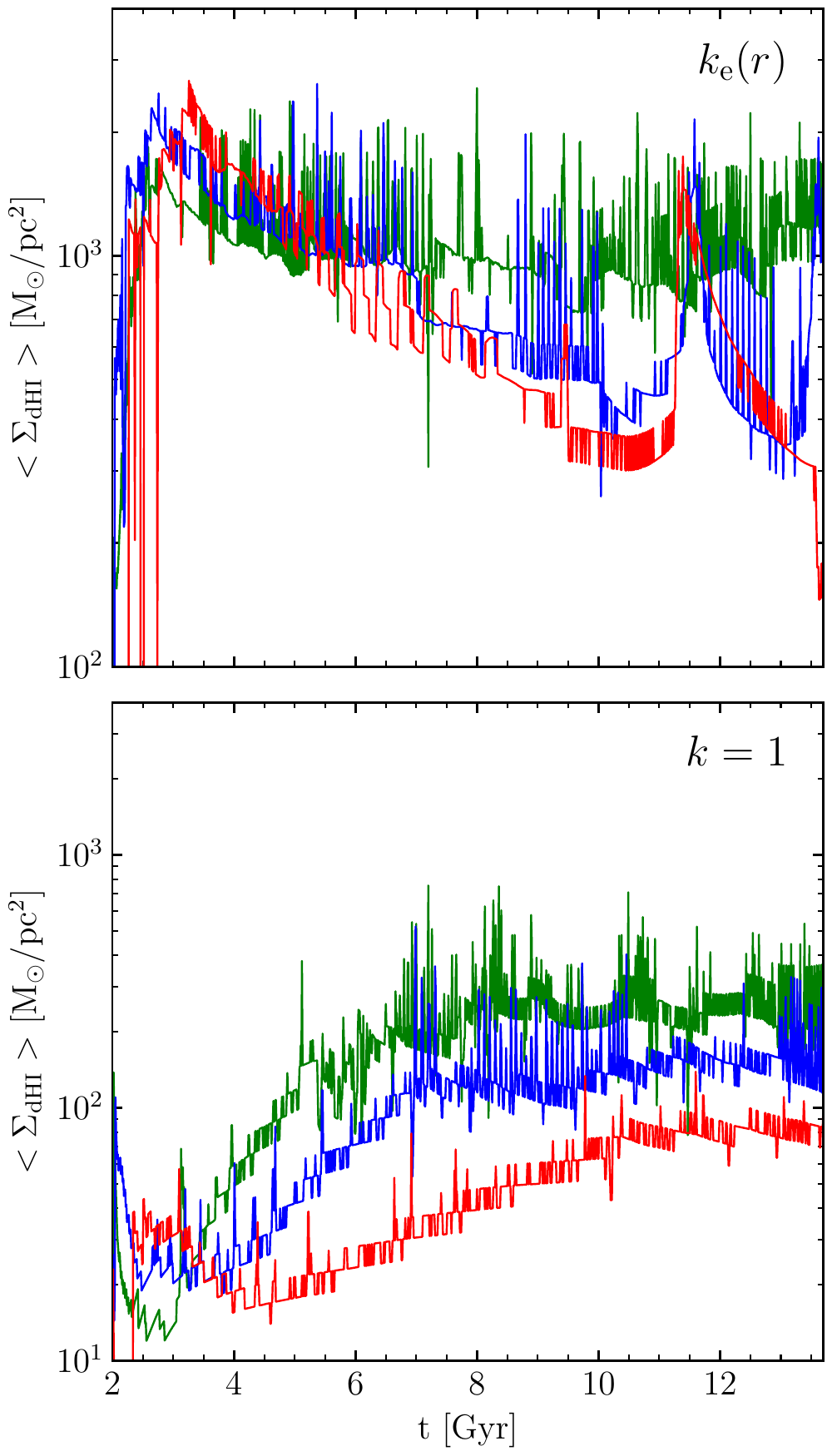}  
  \caption{Left panels: time evolution of the total mass of cold gas
    present in the equatorial disk, for the high-mass (green),
    medium-mass (blue), and low-mass (red) galaxy models, over the
    whole cosmic time spanned by the simulations. The different
    amounts of ordered stellar rotation are indicated by $k=\ke (r)$
    (mild rotation), and by $k=1$ (isotropic rotators). Right panels:
    time evolution of the average gas surface density of the disks,
    defined as $\MdHI/(\pi\RdHI^2)$, for the same models in the left
    panels.}
\label{f5}
\end{figure}

Table \ref{results} lists the final values of the mass $\MdHI$ and
size $\RdHI$ (defined as the truncation radius) of the cold gaseous
disks that form in the equatorial plane: they are defined by
considering the region with the gas temperature
$T\leq\Tc = 5\times 10^5$ K. It is apparent how in each of the three
families, the final mass of the cold disk $\MdHI$ increases with
increasing rotational support of the galaxy, and so does the disk size
$\RdHI$, ranging from a few hundreds pc to a few kpc. The increase of
$\MdHI$ with rotation, at fixed galaxy structure, testifies to the
effect of rotation in enhancing gas cooling over the galaxy body. This
can be clearly seen in the left panels of Figure \ref{f5}, where the
time evolution of $\MdHI$ is shown. In particular, notice how in the
isotropic rotators the epoch of the significant drops of disk mass
happens at later times at decreasing galaxy mass, and how the drops
coincide with the beginning of strong burst in SMBH accretion (bottom
left panels in Figures \ref{f3} and \ref{f4}).

A simple explanation for the increase of $\RdHI$ with the importance
of galactic rotation, can be obtained by considering the equation for
the $z$-component of the angular momentum (per unit mass) $j_z$ of the
gas flows, subjected to the angular momentum injection due to stellar
evolution. Due to the axisymmetry of the simulations, and ignoring for
simplicity viscosity effects of the inflows (at variance with the
evolution of the cold and dense equatorial disks, where
$\alpha$-viscosity is taken into account), it is easy to show that
along the pathlines of fluid elements
\begin{equation}
{D j_z\over Dt}={\dot\rho\over\rho}R\, (\vphib -u_{\varphi}),\quad 
R=r\sin\theta, 
\label{eq:djzdt}
\end{equation}  
where $D/Dt$ is the usual lagrangian derivative, $\vphib$ is the
stellar streaming velocity in Equation (\ref{eq:vphis}), $u_{\varphi}$
is the gas azimuthal velocity, and $R$ the cylindrical radius. The
numerical simulations show that the velocity difference of gas and
stars (in the azimuthal direction) is quite small, so that as a
zeroth-order approximation we can assume $j_z$ is conserved. This
allows us to compute the surfaces of constant $j_z=R\, \vphib(R,z)$
(see Figure \ref{f2}). Under this simplified model, the cooling gas
falls onto the equatorial disk at $\Rin$, where the surfaces of
constant $j_z$ cross the equatorial plane. However, due to the
axysimmetric drift, the rotational velocity of the gas is lower than
the galaxy local circular velocity $\vcirc (R)$, and so the gas will
move inward, ending on a circular orbit of radius $\Rfin$, where
$\Rin\vphib(\Rin,0) =\Rfin\vcirc(\Rfin)$ (see Figure \ref{f1}, right
panel). Figure \ref{f2} shows clearly that the gas falls onto the disk
at significantly larger radii in the isotropic rotators than in the
mildy rotating models. This has interesting consequences: even if the
cold gas mass in isotropic rotators is larger than in models of same
structure but less rotating, yet the much larger disk size implies a
{\it lower} gas surface density; as a consequence the more massive
disks in isotropic rotators are expected to be {\it less} Toomre
unstable than the smaller disks in mildly rotating galaxies of same
structure. These expectations are confirmed by the time evolution of
the mean gas surface density, defined as $\SdHI =\MdHI/(\pi\RdHI^2)$,
shown in the right panels of Figure \ref{f5}, and Table \ref{results}.
We conclude that the larger final masses of the SMBH in isotropic
rotators is a consequence not of more instability events, but of fewer
instabilities each involving larger amounts of mass, due to the larger
$\MdHI$.

Quite naturally, the above findings are also found in the evolution of
star formation, as disk instabilities are related both to SMBH
accretion and star formation.

\subsubsection{Star formation}

\begin{figure*}
  \includegraphics[width=0.48\linewidth, keepaspectratio]{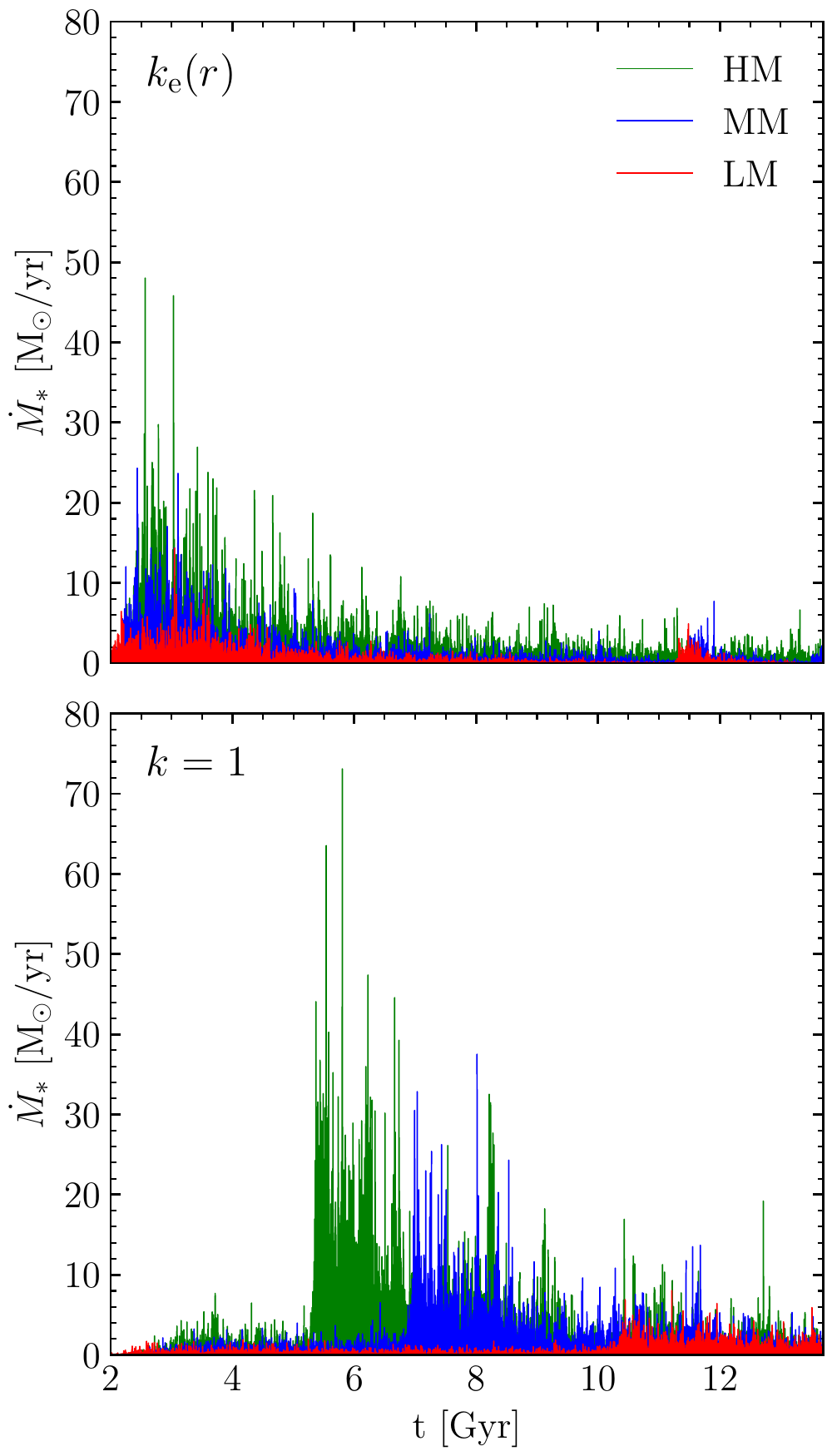}
  \includegraphics[width=0.48\linewidth, keepaspectratio]{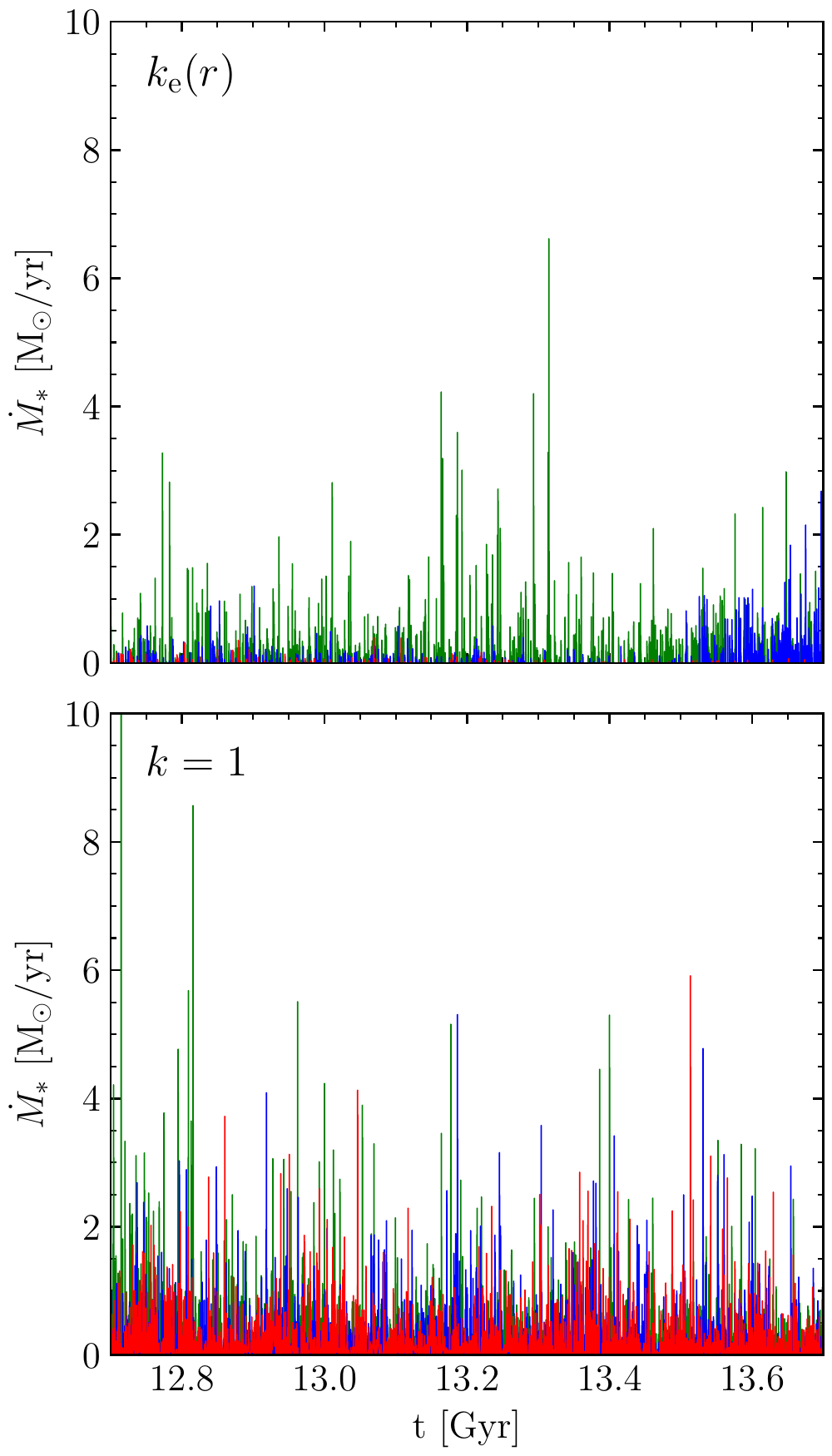}  
  \caption{Time evolution of the star formation rate, for the
    high-mass (green), medium-mass (blue), and low-mass (red) galaxy
    models, over the whole simulated time (left panels), and over the
    last Gyr (right panels). The different amount of galactic rotation
    in the galaxy stellar population is indicated by $k=\ke (r)$ (mild
    rotation), and by $k=1$ (isotropic rotators). }
\label{f6}
\end{figure*}

As anticipated in Section 3, Toomre instabilities in the equatorial
gaseous disk not only lead to mass accretion events on the SMBH, but
also produce local bursts of star formation, as apparent by comparing
Figures \ref{f3} and \ref{f6}, where the SMBH accretion rates
($\dMbh$) and the star formation rates ($\dot M_*$) are shown as a
function of time. The parallel evolution of SMBH accretion (and AGN
activity) and star formation is also visibile in Figure \ref{f4},
where in the right panels we show the cumulative star formation
$\Delta\Mst$ in the galaxy. Again, in the isotropic rotator case, the
less massive galaxies evolve with longer time scales than more massive
systems, as can be seen from the position of the vertical lines in the
bottom-right panel, marking the epoch when half of the total star
formation in each galaxy has been reached.

In the simulations, we assume for simplicity that the newly formed
stars stay on the circular orbit where they form, and we then follow
their evolution, that contributes mass losses, and SNII explosions. At
the end of the simulations, stellar disks of mass
$\Mdstar\simeq 10^8 - 10^9\Msol$, and half-mass radius
$\Rds\simeq 100 - 300$ pc, are present in the equatorial plane
(Columns 5 and 6 in Table \ref{results}); notice that in each family
of models the trend of $\Mdstar$ and $\Rds$ with the galaxy mass and
rotational support nicely follows the trends of the gaseous disks
parameters ($\MdHI$ and $\RdHI$). The stellar disks are significantly
more concentrated ($\Rds <\RdHI$) as a consequence of the density
dependence of the star formation algorithm.  Of course, from the
stellar formation prescription in G19a, star formation is not
necessarily limited to the equatorial gaseous disk; but in the
simulations almost all the star formation takes place in the disk: the
difference between $\Delta\Mst$ and $\Mdstar$ is fully explained by
the star evolution and mass losses in the (top-heavy) secondary star
generations. The inevitable formation of second-generation, metal rich
($\alpha$-enhanced) stellar disks produced by the gas recycled by
stars in the galaxy, is an important prediction of the present models,
that will be discussed in depth in a dedicated paper; notice that
these stellar disks are always corotating with the parent galaxy,
because the intrisic mechanism cannot produce counterrotating disks,
and the material from the circumgalactic medium is assumed to be
accreted on radial orbits.

\subsection{X-ray luminosities and temperatures of the hot ISM coronae}

The last group of quantities characterizing the evolutionary
properties of the hot ISM, are the final values of the total amount of
hot gas $\Mhot$ ($T > \Tc =5\times 10^5$ K), the X-ray luminosity in
the usual $0.3-8$ keV energy band, and measured inside the
obervational aperture of $r_{\rm X}=5\Rec$, and finally the
emission-weighted temperature $\Tx$ measured inside the same aperture
(see Table \ref{results}). An additional quantity useful to check the
mass conservation of the code is the amount of gas $\Mout$ lost at the
last radial grid point ($250$ kpc); in fact for each run we monitored
the mass balance over the whole numerical grid due to the mass sources
and sinks, obtaining an execellent agreement (notice that $\Mout$ and
$\Mhot$ reported in Table \ref{results} cannot be directly compared,
being measured over different volumes).

\begin{figure}
  \includegraphics[width=0.33\linewidth, keepaspectratio]{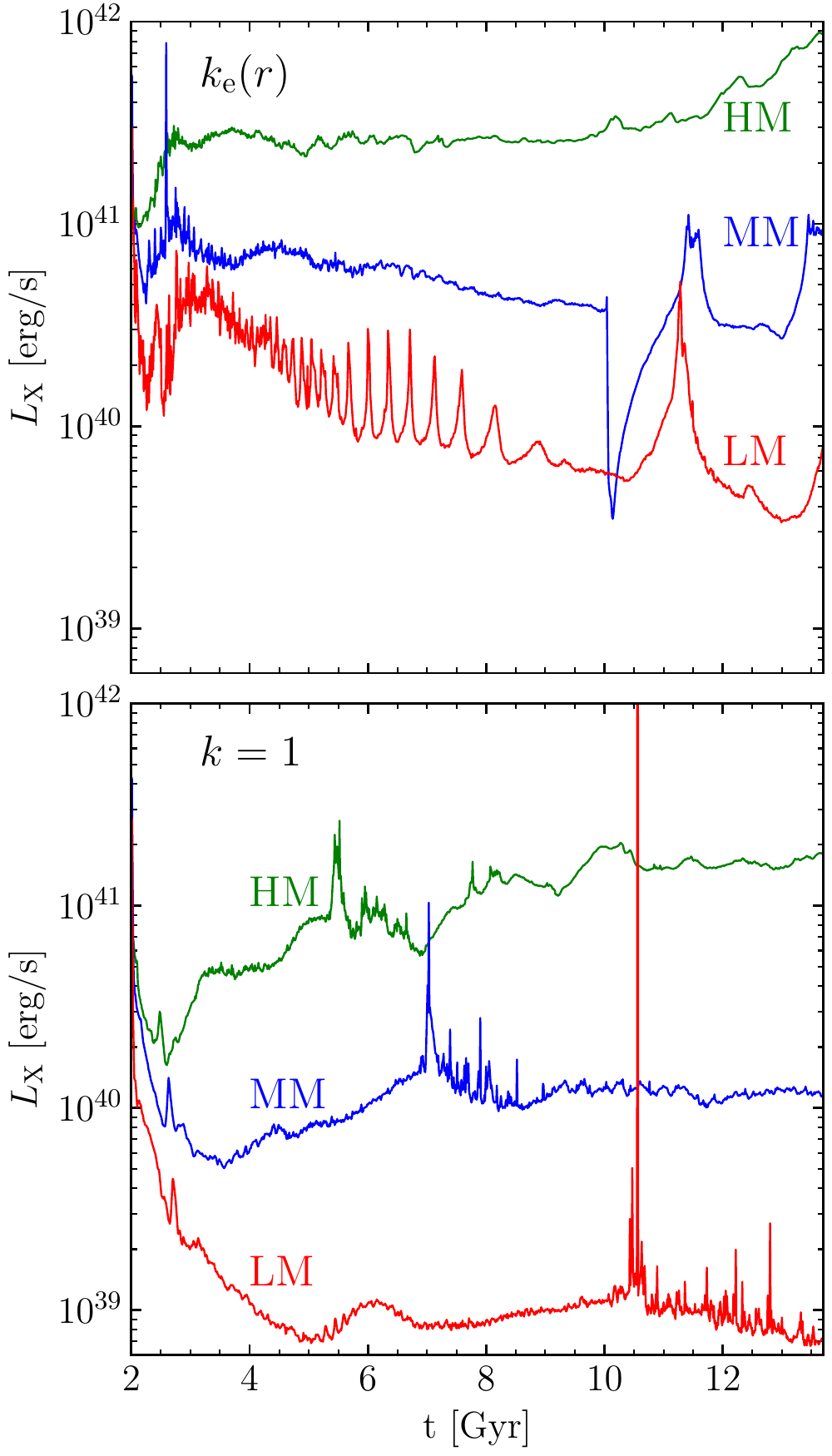}
    \includegraphics[width=0.33\linewidth, keepaspectratio]{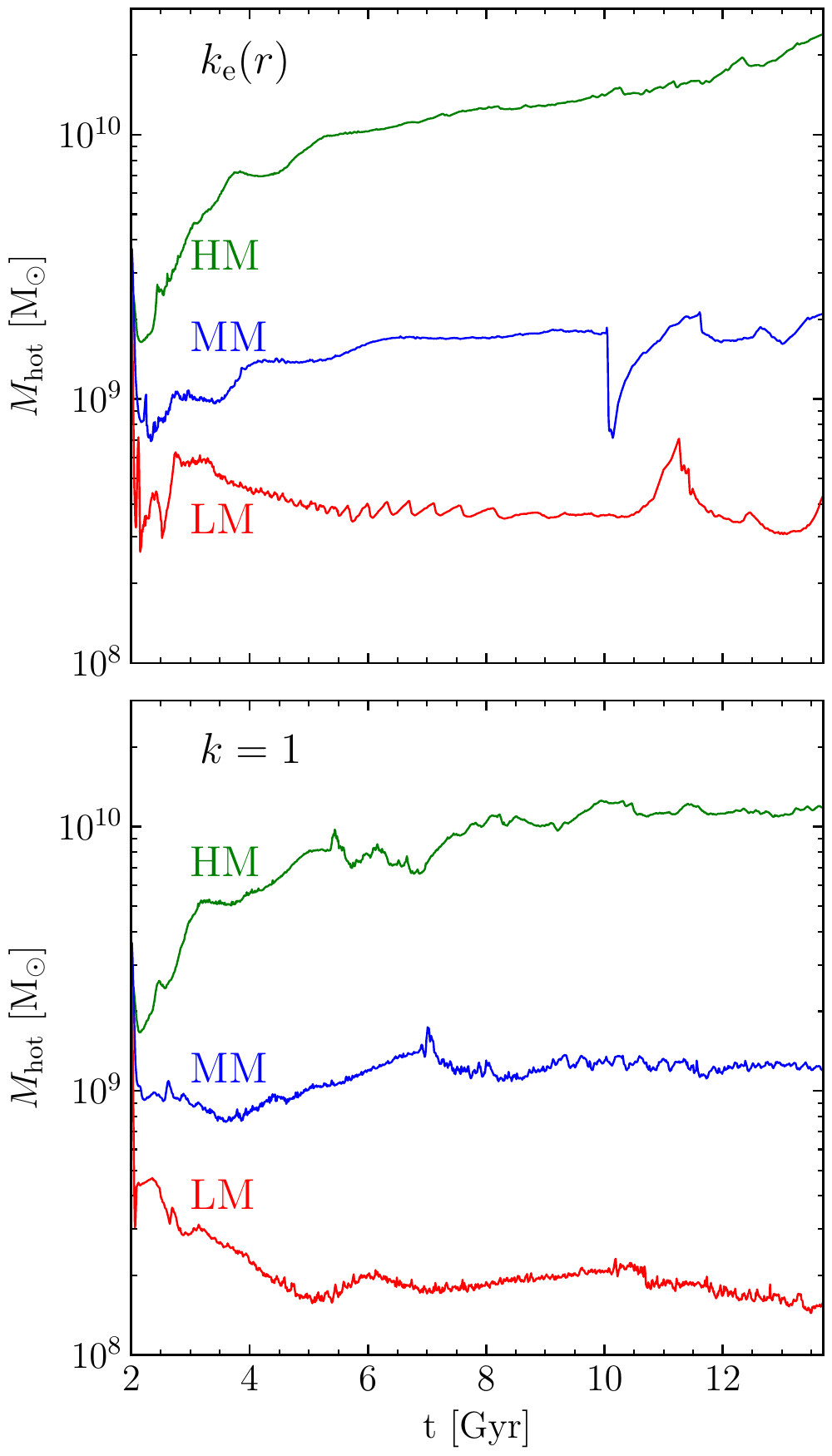}
  \includegraphics[width=0.33\linewidth, keepaspectratio]{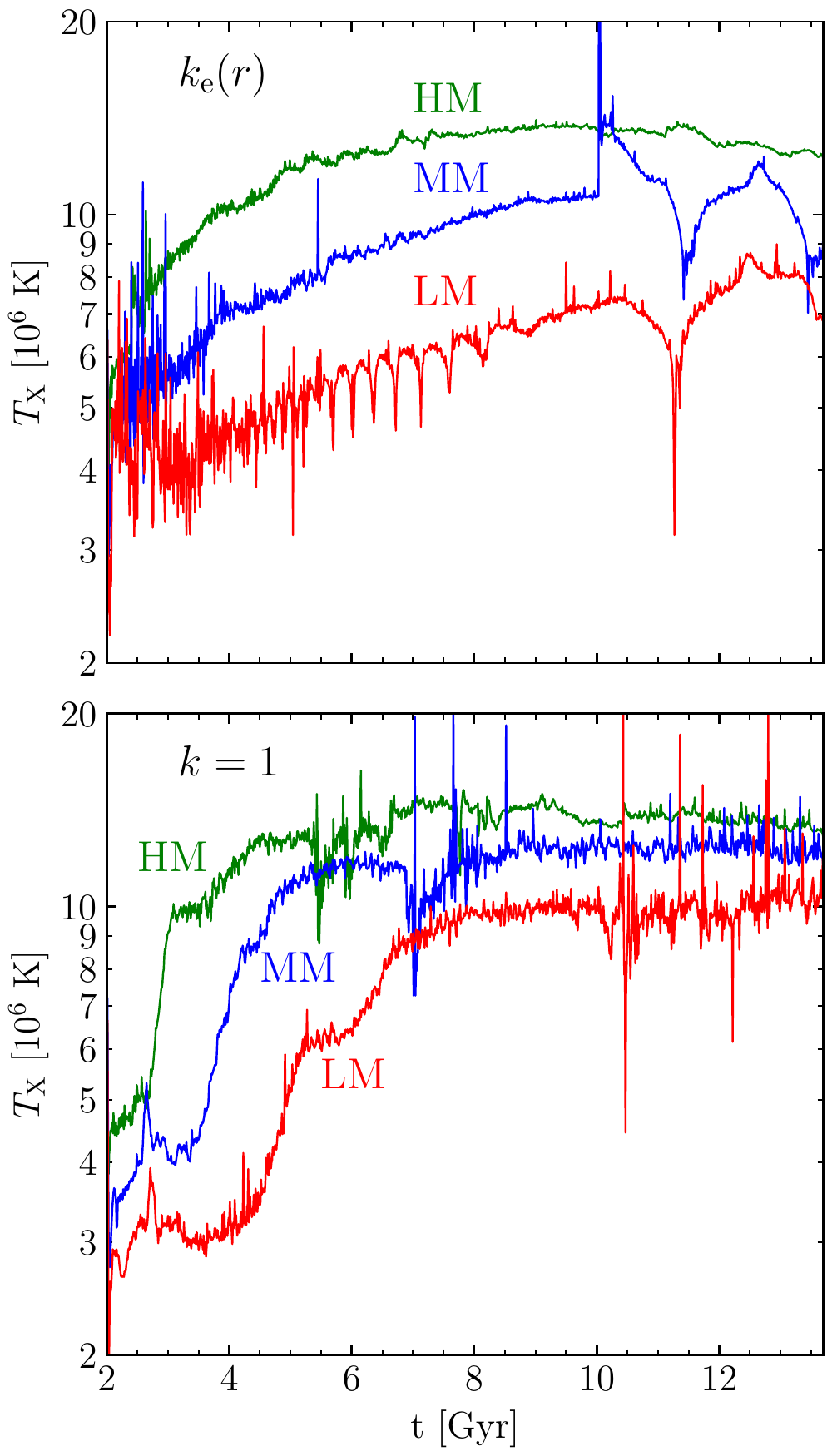}  
  \caption{Left panels: time evolution of the X-ray luminosity $\Lx$
    of the ISM measured in the energy band of $0.3-8$ keV, inside a
    sphere of radius $r_{\rm X}=5\,\Rec$ (see Table \ref{models}), and
    excluding the inner 100 pc to reduce the fluctuations due to AGN
    activity.  Green, blu, and red lines refer to HM, MM, and LM
    models, respectively; note how more ordered galaxy rotation
    (bottom panel) reduces $\Lx$, at fixed galaxy structure.  Central
    panels: time evolution of the mass $\Mhot$ of the hot ISM
    ($T >\Tc = 5\times 10^5$ K) contained within a sphere of radius
    $r_{\rm X} = 5\Rec$. A close parallel between $\Lx$ and $\Mhot$ is
    apparent, with a substantial reduction of $\Mhot$ in the isotropic
    rotators.  Right panels: the emission-weighted X-ray temperature
    of the ISM, over the same volume used for the computation of
    $\Lx$.}
\label{f7}
\end{figure}
\vspace{5truemm}

In Figure \ref{f7} we show the time evolution of $\Mhot$, $\Lx$, and
$\Tx$; reassuringly, the values of $\Lx$ and $\Tx$ agree with those
observed: the final $\Lx$ and $\Tx$ in Table 2 exhibit a range of
  values that compares very well with that reported for the large
  number of ETGs in the $Chandra$ Galaxy Atlas (Kim et al. 2019), and
  in the $Chandra$ sample of Kim and Fabbiano (2015), for galaxies of
  comparable mass. Also, the range of $\Lx$ shown by the models both
  covers most of the observed range, and progressively moves to larger
  values with increasing galaxy mass, as observed. A few trends are
clearly detectable: first, $\Lx$ and $\Tx$ in each family of models
correlate with the total mass of the galaxy, being more massive
galaxies more X-ray luminous and hotter than less massive systems, a
well know manifestation of the underlying Faber-Jackson
relation. Second, $\Tx$ tends to increase with time, while $\Lx$ can
span a range up to two orders of magnitude (for a range of 5 in the
stellar masses in the explored models). Third, for fixed galaxy mass,
less rapidly rotating systems are more X-ray luminous than their
isotropic rotator counterparts. Thus, we confirm that, at each mass,
rotation tends to reduce the X-ray luminosity of ETGs, due to the
tendency of rotating flows to induce gas cooling at relatively large
radii (e.g., Negri et al. 2014a,b; Gaspari et al. 2015).  This
  finding is in accordance with X-ray observations that show flatter
  systems (that are typically more rapidly rotating objects) to have a
  lower $\Lx$ than rounder ones of the same optical luminosity
  (Eskridge et al.  1995; Sarzi et al. 2013; see also Juranova et
  al. 2020). In particular, Sarzi et al., using data from the
  ATLAS$^{3\rm D}$ survey, found fast rotators to have lower $\Lx$ and
  $\Tx$ than slow rotators.  The simulations also predict extended hot
  gas cooling, in rotating systems, and then a larger tendency for
  them to host (large) cold disks. Indeed, gas forming disks in the
  equatorial plane of ETGs (and aligned with the rotation of the
  stars) has been detected from the ionized to the atomic (HI) to the
  molecular (CO) phase, and preferentially in fast rotators (Young et
  al. 2011, Davis et al. 2019, Juranova et al. 2019; see also Babyk et
  al. 2019).  Finally, a paper is in preparation, specifically
  dedicated to a thorough analysis of the X-ray properties of the
  models, including the radial profiles of their X-ray surface
  brightness, and of their luminosity-weighted projected temperature,
  to be compared with those typically observed.

Also note the solution to the classical ``cooling
flow problem'' indicated by our numerical solutions. While some (metal
enriched) gas falls to the center, as revealed by $\Delta\Mbh$,
$\MdHI$, and $\Mdstar$, $\approx 30$ times more gas ($\Mout$) is
expelled by feedback, as reported in Column 9 of Table \ref{results}.
In general, rotating models (with the exception of HM family) tend to
eject more mass as the rotational support of the galaxy increases,
because rotation not only increases the tendency for gas cooling, but
also unbinds gas at large radii (see e.g. Cotti and Pellegrini 1996,
Posacki et al. 2014, Negri et al. 2014b); thus the net effect of
substantial rotation is to produce more cold gas and less hot ISM,
leading to an X-ray underluminosity and lower hot gas temperatures.

\section{Discussion and conclusions}

In this paper we presented a first, systematic exploration of the hot
gas evolution for a set of realistic high resolution models of massive
ETGs with central SMBHs. The exploration was conducted with the latest
version of the high-resolution 2D hydrodynamical code MACER. The
innermost grid point was placed at 25 pc from the center, the
outermost at 250 kpc, and the flow evolution was followed at high
temporal resolution over the cosmological time span of 12 Gyr.  A few,
time-expensive test simulations, were also conducted with a much
higher spatial resolution, with the first active grid point placed at
2.5 pc from the SMBH.  The initial stellar mass of the galaxy models
is in the range $1.5\times 10^{11} <\Mst/\Msol <7.8\times 10^{11}$,
and has the E3 shapes when observed edge-on. The stellar density
distribution, and the DM halo associated with the galaxies, are
modeled by two-component ellipsoidal Jaffe profiles (JJe models,
CMPZ21), providing a very good approximation over a large radial range
of the de Vaucouleurs and the NFW profiles, respectively; a
group/cluster quasi-isothermal DM halo with a flat rotation curve in
the outer regions is also considered. The internal dynamics of the
galaxies is obtained by solving the Jeans equations, and for each
model we explore the non-rotating case (when the galaxy flattening is
fully supported by tangential velocity dispersion), the isotropic
rotator (when galaxy flattening is fully supported by ordered
rotation), and an intermediate case with exponentially declining
ordered rotation, obtained from a spatially dependent Satoh
decomposition.  Mass sources are represented by mass losses from stars
(red giants, AGB stars, and SNIa/SNII explosions, computed following
the prescriptions of stellar evolution), and by a time-dependent
cosmologically motivated mass accretion rate from the group/cluster
ambient, imposed at the outer boundary of the numerical grid.  In
rotating models the stellar mass losses are injected in the ISM
following the galaxy ordered velocity field, and the cooling gas
collapses on to a rotating gaseous disk in the equatorial plane. Tqhe
ISM is heated by thermalization of the kinetic energy of SNe
explosions, and stellar motions; gas cooling is implemented as in our
previous version of MACER (G19a); the production and circulation of
metals, and the formation/destruction of dust, are also considered
following G20.  Two different channels are considered for star
formation: the classical one based on the cooling and and the Jeans
collapse times of the ISM, and a second based on the assumption that
the rotating gaseous disk self-regulates due to Toomre instabilities
around a value of $Q\simeq 1$. These instabilites lead to bursts of
star formation, the formation of a central rotating stellar disk,
outward angular momentum transport and inward mass transport (in
addition to the effects of standard $\alpha$-viscosity, also
considered in the simulations), and finally to SMBH accretion and AGN
feedback.

As a first improvement over our previous simulations, we consider the
secular evolution of the galaxy gravitational field due to mass losses
of stars (in addition to the changes of the gravitational field due to
the mass growth of the SMBH, already considered in our previous
studies); we also implemented the associated changes of the velocity
dispersion and rotational fields of the stars. In rotating models, the
effects of the time-evolving gravitational field of the equatorial
stellar disk on the gas flows, are also taken into account. As a
second important improvement we now model the UV heating effects of
the massive, young stars in the stellar disk, in addition to the disk
SNII feedback.  The third set of improvements concerns the treatment
of AGN feedback. In particular we adopt a higher maximum wind
efficiency $\epswM$ in the cold-accretion mode, and a smoother
transition of $\epsw$ between cold and hot accretion regimes.

The main results can be summarized as follows. In general, we confirm
the picture that the evolution of the ISM undergoes recurrent cycles,
during which the gas cools, falls towards the central galactic
regions, and -- if it possesses angular momentum -- accumulates in a
central disk; there, it becomes over-dense and self-gravitating, until
in the disk the Toomre instability sets in, allowing for star
formation and mass inflow from the disk towards the SMBH.  The
errupting SMBH then ejects much of the inflowing material back into
the ISM.  Thus, with a short delay (of the order of the orbital period
of the circumnuclear disk), star formation is followed by accretion of
disk material onto the SMBH.  An AGN burst is then triggered, and the
energy output from the galactic centre, in the form of radiation and
winds, modifies the hydrodynamics of the ISM throughout the host
galaxy (which is known as the AGN feedback).  The biconical AGN winds
cause the ejection of gas into the polar regions, but also the other
galactic regions are affected more or less directly by the propagation
of shock waves, with the consequent alternate compression and
rarefaction.  After a starburst, the massive stars can also feed
energy back to the ISM via SNII explosions; this impacts mostly the
region around where star formation occurs (over a lengthscale of
$\sim 1$ kpc). Most of the SNII events occur within the cold rotating
(and dusty) disk, but in some models we allow for 40\% of the SNII to
arise from runaway stars which have typically travelled 100--300 pc
away from their birthplaces. The new stars in the central disk form
with a top heavy mass function as found in the MW and M31 (see also
Goodman and Tan 2004). They are embedded in a dusty cool gas envelope
which will have notable IR emission properties (see e.g. G19b), in
agreement with observations.

More in detail, we focused on three specific properties of the model
evolution, considering both the effects of the galaxy mass, and of the
degree of internal rotation. 

For what concerns SMBH accretion, we found (not suprisingly) that
$\Delta\Mbh$ increases with galaxy mass, but more than linearly with
the mass sources, i.e., SMBHs in massive galaxies accrete more
efficiently than SMBHs in galaxies of lower mass, a natural
consequence of the scaling with galaxy mass of heating sources and the
depth of the galaxy potential well, with the SMBH mass accretion (and
AGN feedback) peaking earlier in the high mass systems. Moreover, at
fixed galaxy mass the more rapidly rotating galaxies accrete more
material on to their central SMBH.  This is due to the fact that a
stronger rotation tends to favour large scale instabilities and gas
cooling, leading to stronger inflows, and the formation of more
massive and extended gaseous disks.  The larger $\Delta\Mbh$ of fast
rotators is due to fewer instability events in the disk, characterized
though by significantly larger mass accretion. In fact, accretion
reaches systematically higher $\dMbh$ in high mass models and in
models with substantial internal rotation. Important accretion
episodes begin almost immediately in the mildy rotating galaxies,
while the first massive accretion episodes in the isotropic rotators
start at quite late times, with the epoch of the first important event
increasing at decreasing galaxy mass. It is intriguing to speculate
that these trends may help to explain the empirical observation that
the activity of lower mass Seyfert galaxies peaks at later epochs than
do higher mass Quasars. Overall, the results in this Section confirm
that AGN feedback is efficient to maintain SMBHs masses in the present
universe small, when compared to the available gas that could be
accreted with unstopped cooling flows.

For what concerns the formation of the equatorial gaseous disk, its
instabilites, and the associated star formation, we confirmed that gas
cooling, even in presence of moderate rotational support of the
stellar component, produces cold gaseous disks in the equatorial
plane, with present day masses in the range $10^8\Msol -10^9\Msol$,
sizes ranging from a fraction of kpc to a few kpc, and surface
densities of $\approx 10^2\Msol/{\rm pc}^2$; masses and disk sizes
increase for increasing galaxy mass and amount of rotational support.
Interestingly, even if the mass of the cold disks in isotropic
rotators is larger than in models of same structure but less rapidly
rotating (due to the well known enancement of cooling efficiency in
rotating models), yet the much larger size implies a lower gas surface
density, so that the more massive disks in isotropic rotators are in
general {\it less} Toomre unstable than the smaller disks in moderatly
rotating galaxies of same structure.  An interesting consequence of
this behavior is that the larger final masses of the SMBH in isotropic
rotators are a consequence not of more instability events, but of
fewer instabilities each involving larger amounts of mass, due to the
larger values of $\MdHI$.  As instabilities in the gaseous disk, not
only lead to mass accretion events on the central SMBH, but also
produce local bursts of star formation, we also found at the end of
the simulations, stellar disks of mass
$\Mdstar\simeq 10^8\Msol - 10^9\Msol$, and half-mass radii
$\Rds\simeq 100 {\rm pc}- 300$ pc, in the galaxy equatorial plane; in
each family of models the dependence of $\Mdstar$ and $\Rds$ on the
amount of galaxy mass and rotational support nicely follows the trends
of the gaseous disks properties $\MdHI$ and $\RdHI$.  Moreover, in the
isotropic rotator case, the less massive galaxies evolve with longer
time scales than the more massive systems.

Finally, for what concerns the X-ray properties of the hot gas, in our
systematic exploration of parameter space, the values of $\Lx$ (the
X-ray luminosity inside $5\Rec$ and in the energy band of 0.3-8 keV)
and of $\Tx$ (the associated emission-weighted temperature over the
same volume) are in the observed range, with more massive galaxies
hosting more luminous gaseous halos.  In each mass range, the
isotropic rotators are found at a lower luminosity than models of
similar structure but less rapidly rotating, confirming that rotation
tends to reduce the X-ray luminosity of galaxies, due to the strong
tendency of rotating flows to induce gas cooling.  We also confirm the
strong sensitivity of X-ray luminosity on the galaxy mass, with $\Lx$
spanning a range up to two orders of magnitude, for a range of a
factor of 5 in the stellar masses.

There are numerous observational checks possible to determine if we
have adequately modelled the evolution of gaseous halos of massive
galaxies, and we list here some of them. Do the final hot X-ray
properties agree with observations in terms not only of intergrated
properties, but also on detailed radial profiles of $\Sigma_{\rm X}$
and $\Tx$? Does the amount and metallicity of the expelled gas
correspond to the observed CGM?  Do the predicted circumnuclear gas
and stellar disks exist in the real world? Of course mergers, which we
neglect, would tend to disrupt and disperse this component.  Do the
outflowing winds seen in AGN have the high metal content -- in
particular the $\alpha$-enhanced abundances -- predicted by our models
as a consequence of top-heavy star formation in the central disk?
What is the effect of a nuclear jet on the galaxy evolution? What new
phenomena are associated with genuine 3D hydrodynamics?  Further
papers in this series will address some of these questions.

\acknowledgements We thank the several scientists who have helped us
in this work, including Ralf Bender, Michele Cappellari, Ena Choi,
Bruce Draine, John Kormendy, Raffaella Morganti, Thorsten Naab,
Tom Oosterloo, and Feng Yuan.  We acknowledge computing
resources from Columbia University’s Shared Research Computing
Facility project, which is supported by NIH Research Facility
Improvement Grant 1G20RR030893-01, and associated funds from the New
York State Empire State Development, Division of Science Technology
and Innovation (NYSTAR) Contract C090171, both awarded April 15,
2010. We are also pleased to acknowledge that the work reported on in
this paper was substantially performed using the Princeton Research
Computing resources at Princeton University which is consortium of
groups including the Princeton Institute for Computational Science and
Engineering and the Princeton University Office of Information
Technology’s Research Computing department.

\newpage
\onecolumngrid
\appendix

\section{Stellar velocity dispersions}
\label{app:ABCDEF}

The solution of the Jeans equations for the galaxy stellar component
(excluding for simplicity the contribution of the group/cluster
quasi-isothermal DM halo and of the time-dependent equatorial stellar
disks, see Section 2.1) can be written as
\begin{equation}
      \sigmas^2=\sigmaBH^2+\sigmag^2,\qquad\Dels=\DeltaBH+\Deltag,
\end{equation}
where $\sigmaBH$ and $\sigmag$ represent the contribution of the
central SMBH and of the galaxy potential to the radial and vertical
components of the velocity dispersion tensor, and similarly for the
quantity $\Dels=\vphib^2+\sigmaphi^2 -\sigmas^2$. For the considered
models, in the special case of a spherically symmetric {\it total}
(stars plus DM) density distribution, the general solutions (see
CMPZ21) reduce to
\begin{equation}
      \rhost\sigmaBH^2= 
      {G\Mst^2\mu\over 4\pi\rst^4}\left[A(s)+\etas B(s)+\etas C(s)s^2\sin^2\theta\right],
     \quad
  \rhost\DeltaBH=
  {G\Mst^2\mu\over 2\pi\rst^4}\etas C(s)s^2\sin^2\theta,
 \label{eq:rhos_sigsDBH}
\end{equation}
\begin{equation}
     \rhost\sigmag^2= 
      {G\Mst^2\MR\over 4\pi\rst^4}\left[D(s,\xi)+\etas E(s,\xi)+\etas F(s,\xi)s^2\sin^2\theta\right], 
\qquad
\rhost\Deltag=
{G\Mst^2\MR\over 2\pi\rst^4}\etas F(s,\xi)s^2\sin^2\theta,
\label{eq:rhos_sigsDg}
\end{equation}
where $s=r/\rst$ and 
\begin{equation}
A(s)={12s^3 + 6s^2 - 2s + 1\over 3s^3(1+s)}+4\ln{s\over 1+s},\quad
B(s)={24 s^4 +36 s^3 + 8 s^2 - 2 s -1\over 3s^3(1+s)^2}+8\ln {s\over1+s},
\end{equation}
\begin{equation}
C(s)=-{180 s^6 +270 s^5 + 60 s^4 -15 s^3 + 6 s^2 -3 s - 4\over
  10s^5(1+s)^2} -18\ln {s\over 1+s}.
\end{equation}
\begin{equation}
  D(s,\xi)=-{3\xi^2-\xi-1\over \xi^2(\xi-1)(1+s)}
                 - {(3\xi+2)s-\xi\over 2\xi^2s^2(1+s)}
                 -{1\over \xi^3(\xi-1)^2}\ln {s\over \xi+s}
                 -{3\xi-4\over (\xi-1)^2}\ln{s\over 1+s},
\end{equation}
\begin{eqnarray}
E(s,\xi)&=&-\frac{2(3\xi^3-6\xi^2+2\xi-1)s + 9\xi^3-18\xi^2+9\xi-4}{2\xi^2(\xi-1)^2(1+s)^2}
            -\frac{2(\xi-1)s+\xi}{2\xi^2s^2(1+s)^2}\cr
            &+&\,\frac{3\xi-1}{\xi^3(\xi-1)^3}\ln\frac{s}{\xi+s} 
-\frac{3\xi^2-9\xi+8}{(\xi-1)^3}\ln\frac{s}{1+s}, 
\end{eqnarray}
\begin{eqnarray}
F(s,\xi)&=&
\frac{2(5\xi^5-8\xi^4+\xi^3+\xi^2+\xi-1)s
            +15\xi^5-24\xi^4+3\xi^3+3\xi^2+5\xi-4}{\xi^4(\xi-1)^2(1+s)^2}\cr
            &+&\,\frac{4(5\xi^3+2\xi^2-3)s^3 -\xi(5\xi^2+2\xi-6)s^2 +2\xi^2(\xi-2)s+3\xi^3}{6\xi^4s^4 (1+s)^2}\cr
            &-&\frac{2(2\xi-1)}{\xi^5(\xi-1)^3}\ln\frac{s}{\xi +s} +\frac{2(5\xi^2-13\xi+9)}{(\xi-1)^3}\ln\frac{s}{1+s}.
\end{eqnarray}
Finally, the functions $D$, $E$, and $F$ need a separate treatment for
the special case $\xi=1$, when
\begin{equation}
D(s,1)=-{(6s^2 + 6s -1)(2s +1)\over 2s^2 (1+s)^2}  - 6\ln{s\over 1+s},\quad
E(s,1)=-{12s^4 + 30s^3 + 22s^2 + 3s + 3\over 6s^2 (1+s)^3} -2\ln{s\over 1+s},
\end{equation}
\begin{equation}
F(s,1)={60s^6 + 150s^5 + 110s^4 + 15s^3 - 3s^2 + s + 3\over 6s^4(1+s)^3}+10\ln{s\over 1+s}.
\end{equation}

\section{Gravitational effects of stellar mass losses and SMBH growth}
\label{app:MassLoss}

One of the useful features of the adopted analytical models for
galaxies is the possibility to easily implement in the hydrodynamical
code the secular changes of the gravitational field of the galaxy and
of the stellar velocity dispersion and rotational fields of the stars
due to the mass growth of the central SMBH and to the reduction of the
stellar mass due to the stellar mass losses. Moreover, as described in
Sections 2.1 and 3, we also consider the effects of the time
independent gravitational field of a group/cluster DM halo, and the
time dependent gravitational field of the stellar equatorial disk
produced by the rotating cooling gas: hower, for simplicity we neglect
the effects of these two gravitational fields on the velocity fields
of stars.

The stellar mass losses (stellar winds plus SNIa explosions) produce a
mass source term $\dot{\rho}=\alpha(t)\rhost$ in the hydrodynamical
equations, where the function $\alpha(t)$ is prescribed by stellar
evolution (see e.g. Pellegrini 2012; Ciotti \& Ostriker 2012 for
details). We define the {\it mass reduction factor}
\begin{equation}
  f(t)=1-\epsilon(t), 
  \qquad 
  \epsilon(t)\equiv\int_{2{\rm Gyr}}^t\alpha(\tau)d\tau, 
\end{equation}
so that 
\begin{equation}
      \Mst(t)=f(t)\Mst, 
      \quad
      \rhost(t)=f(t)\rhost, 
      \quad
      \phis(t)=f(t)\phis, 
\end{equation}
and in the following all quantities independent of time refer to the
initial time of the simulations (when as usual the stellar population
is assumed to be 2 Gyr old). In particular, in equation above $\phis$
is the potential at the beginning of the simulations of the
ellipsoidal Jaffe stellar distribution in Equation (\ref{eq:rhog}),
obtained for simplicity by homeoidal expansion
\begin{equation}
      \phis={G\Mst\over\rst}
      \left[
      \tphisz(s) 
      +\etas\tphisu(s) 
      +\etas\tphisd(s)s^2\sin^2\theta 
      \right], 
      \quad
      \tphisi=
        \begin{cases} 
        \displaystyle{\ln{s\over 1+s}},\qquad  (i=0), \\
        \displaystyle{{s^2+2s+4\over 3s^2(1 +s)} + {1\over3}\ln{s\over1+s}-{4\ln (1+s)\over 3s^3}},\quad  (i=1),\\
        \displaystyle{-{s+2\over s^4(1+s)} + {2\ln (1+s)\over s^5}},\quad (i=2),
        \end{cases}
\end{equation}
(see Equation 19 in CMPZ21, with $\xi=\MR=1$ and $\etag=\etas$
therein).  Notice that in terms of the initial quantities,
\begin{equation}
\MR(t)\equiv{\Mgal(t)\over\Mst(t)}={\MR-\epsilon(t)\over f(t)},\quad
\mu(t)\equiv{\Mbh(t)\over\Mst(t)}={\mu\over f(t)}{\Mbh(t)\over\Mbh}.
\end{equation}
The total gravitational potential experienced by the gas flows can be
written
\begin{equation}
      \phiT(t)=\phigal + \phih + {\Mbh(t)\over\Mbh}\phiBH + \phids(t) -\epsilon(t)\phis,
\label{eq:phitt}
\end{equation}
where $\phigal$, $\phih$, $\phiBH$ and $\phids(t)$ are given
respectively by Equations (\ref{eq:phigal}), (\ref{eq:phih}),
(\ref{eq:phiBH}) and (\ref{eq:phidisk}).

Finally, we obtain the expression for the time dependence of the
vertical (and radial) velocity dispersion $\sigmas$ and of the
function $\Deltas$ needed in Equation (\ref{eq:vphis}) to determine
the azimuthal velocity dispersion and the streaming velocity of
stars. From the dependence of the Jeans equations on the total
potential, and from the considerations above, it is easy to show that 
\begin{equation}
\sigmas^2(t)=\sigmag^2 + {\Mbh(t)\over \Mbh}\,\sigmaBH^2 
-\epsilon(t)\sigmass^2, \quad
\Deltas(t)=\Deltag + {\Mbh(t)\over \Mbh}\,\DeltaBH 
-\epsilon(t)\Deltass,
\end{equation}
where the time independent quantities $\sigmag$, $\Deltag$, $\sigmaBH$
and $\DeltaBH$ are obtained from Equations (A1)-(A2) by using Equation
(\ref{eq:rhos}).  $\sigmass$ and $\Deltass$ describe the
self-contribution of the stellar distribution. From Equations (39) and
(41) in CMPZ21 one obtains
\begin{equation}
\rhost\sigmass^2 = {G\Mst^2\over 
  4\pi\rst^4}\left[ D(s,1)+\etas X(s) + \etas Y(s)
  s^2\sin^2\theta\right], \quad
 \rhost\Deltass ={G\Mst^2\over 2\pi\rst^4}\etas Z(s) s^2\sin^2\theta,
\end{equation}
where
\begin{eqnarray}
  X(s)&=& {86s^6+185s^5+101s^4-s^3-4s^2-4\over 5s^4(1+s)^3} -{34\over 5}\ln {s\over 1+s}\\
        &+&2\left[{60s^5+30s^4-10s^3+5s^2-3s+2\over 5s^5(1+s)}+12\ln{s\over 1+s}\right]\ln(1+s)-24\Hcsi(1,s), 
\end{eqnarray}
\begin{eqnarray}
Y(s)&=&-{4572s^8+10170s^7+6002s^6+198s^5-90s^4+44s^3-60s^2-15s-90\over63s^6(1+s)^3}+{52\over
        7}\ln {s\over 1+s}\\
    &-&2\left[{840s^7+420s^6-140s^5+70s^4-42s^3+28s^2-20s+15\over
        21s^7(1+s)} +  40\ln {s\over 1+s}\right]\ln(1+s)+80\Hcsi(1,s),
\end{eqnarray}
\begin{eqnarray}
Z(s)&=&-2{2286s^7+2799s^6+202s^5-103s^4+58s^3-36s^2+6s+18\over
        63s^6(1+s)^2}+{52\over 7}\ln{s\over 1+s}\\
    &-&2\left[{840s^8+1260s^7+280s^6-70s^5+28s^4-14s^3+8s^2-5s-6\over 21s^7(1+s)^2}  + 40\ln{s\over 1+ s}\right]\ln(1+s)+80\Hcsi(1,s).
\end{eqnarray}
and the function $\Hcsi(\xi,s)$ is defined in Equation (83) of CMPZ21.

\bigskip\bigskip

\twocolumngrid

\end{document}